# What Really Drives Thermopower: Specific Heat or Entropy as the Unifying Principle Across Magnetic, Superconducting, and Nanoscale Systems

Morteza Jazandari,[a] Jahanfar Abouie,[a*] and Daryoosh Vashaee[b,c*]

[a] Department of Physics, Institute for Advanced Studies in Basic Sciences (IASBS), Zanjan 45137-66731, Iran
[b] Department of Electrical and Computer Engineering, North Carolina State University, Raleigh, NC 27606, USA
[c] Department of Materials Science and Engineering, North Carolina State University, Raleigh, NC 27606, USA

Thermopower, a key parameter in thermoelectric performance, is often linked to either specific heat or entropy, yet the fundamental quantity that governs it has remained elusive. In this work, we present a unified theoretical framework that identifies entropy per carrier, not specific heat, as the universal driver of thermopower across both closed and open systems. Using thermodynamic identities and the Onsager–Kelvin relation, we show that thermopower is universally proportional to entropy per carrier, while its apparent proportionality to specific heat arises only in systems where the specific heat follows a continuous power-law temperature dependence. To extend this framework to magnetic systems, we derive a general expression for magnon-drag thermopower that holds in both Newtonian (massive, parabolic) and relativistic (massless, linear) magnon regimes. In particular, we reformulate the momentum balance using a relativistic energy-momentum tensor, resolving conceptual inconsistencies in prior models that relied on ill-defined magnon masses in antiferromagnets. Our framework is further illustrated through three representative systems: (i) magnetic materials, where magnon and paramagnon entropy sustain thermopower across $T_C$ and $T_N$; (ii) superconducting Nb, where anomalous thermopower emerges from entropy carried by Bogoliubov quasiparticles near $T_C$; and (iii) a single-molecule junction, where entropy from occupation-number fluctuations governs thermopower in an open quantum system. We validate our unifying principle by comparing it with experimental data: thermopower measurements of superconducting niobium reveal the role of quasiparticle entropy near the critical temperature, and literature-reported specific heat data from a wide range of ferromagnetic and antiferromagnetic materials demonstrate consistent entropy-based scaling across magnetic transitions. These case studies demonstrate that entropy provides a unifying thermodynamic foundation for understanding thermopower across quantum, classical, magnetic, and nanoscale systems.

## I. INTRODUCTION

Thermopower, or the Seebeck coefficient, measures the voltage generated by a material due to a temperature difference. Over the years, various theoretical frameworks have been developed to quantify and predict thermopower behavior in different materials and conditions.

When discussing thermopower, we consider a system influenced by external fields, including a temperature gradient $\nabla T$ and a time-dependent electric field $\nabla \phi$. For the condition, electric current $\boldsymbol{J} = 0$, thermopower can be expressed as a function of wave vector $\boldsymbol{k}$ and frequency $\omega$ in its most general form as follows:

$$\alpha(\boldsymbol{k}, \omega) = \frac{\nabla \phi}{\nabla T} \{\boldsymbol{k}, \omega\}, \qquad (1)$$

where $\{\boldsymbol{k}, \omega\}$ denotes the two arguments of the binary function $\nabla \phi / \nabla T$.

Deriving the general form of the thermopower equation is exceedingly challenging. Instead, researchers typically formulate the thermopower equation for each system using approximations tailored to its specific characteristics. Thermopower is generally analyzed and quantified under two regimes: AC and DC. By considering the limiting values of wave vectors and frequencies, various methods have been employed to determine thermopower behavior effectively.

One common method, known as the fast DC limit, first considers the wave vector $\boldsymbol{k} \to 0$ and then the frequency $\omega \to 0$. In this limit, the thermopower is given by the Kubo formula: $\alpha_{\text{Kubo}} = \lim_{\omega \to 0} \lim_{\boldsymbol{k} \to 0} \alpha(\boldsymbol{k}, \omega)$. This limit is used for DC transport and is applicable for dissipative systems such as correlated electron systems [1,2]. However, solving the Kubo formula for interacting systems is computationally complex, and this method is primarily applied to non-interacting systems. While the Kubo formalism is challenging to solve analytically, it is predominantly computed numerically. The AC limit of thermopower $\alpha^*$, also known as the high-frequency method, is optimal for systems in which the relaxation time is overshadowed by exceedingly large frequencies. The many-body information contained in the Kubo formula is largely captured by $\alpha^*$ at high frequencies [1,2]. This method provides a practical formula that requires considerably less computational effort. Another formula for the DC regime is called the Mott formula $\alpha_{\text{Mott}}$ which is a consequence of the Kubo formula in the DC limit under the assumption of weak scattering and is suitable for metals, insulators, and intrinsic semiconductor systems [2,3]. The high-temperature limit of thermopower (Kubo formula) for narrowband systems is known as the Heikes-Mott (HM) formula $\alpha_{\text{HM}}$ [2].

Ultimately, the Kelvin formula is obtained under conditions where, according to Onsager's prescription, the frequency and then the wave vector of the electric field tend

---

to zero. $\alpha_{\text{Kelvin}} = \lim_{k\to 0} \lim_{\omega\to 0} \alpha(k,\omega)$ is the slow DC limit of $\alpha(k,\omega)$ [4,5]. In this case, the periodic boundary condition can be considered equivalent to the open boundary condition (as considered in experiments), and the thermodynamic equilibrium state can be defined from the transport equations to calculate thermopower. $\alpha_{\text{Kelvin}}$ is generally proportional to the derivative of entropy $S$ with respect to the particle number $N$ at constant temperature $T$ and volume $V$. Compared to methods such as the Kubo method and dynamical mean-field theory, the Kelvin method offers distinct advantages. It avoids computational challenges like the analytical continuation required in the Kubo formula for calculating dynamic response functions such as electrical conductivity. Furthermore, it explicitly expresses thermopower in terms of thermodynamic quantities, providing a direct framework for optimizing thermoelectric performance in experiments. Although each of these formulas is suited to specific regimes, they become equivalent under certain conditions. For example, in electronic systems at high temperatures - where $k_BT$ dominates over system's relevant energy scales - the Kelvin and Heikes-Mott thermopower expressions ($\alpha_{\text{Kelvin}}$ and $\alpha_{\text{HM}}$) converge when the ratio of chemical potential to temperature ($\mu/T$) remains temperature-independent. This equivalence holds across various systems; for instance, it has been demonstrated by Arsenault et al. [1] for doped Mott insulators on the fcc lattice, and similarly applies to the single-molecule junction discussed in Section VII. However, at lower temperatures ($k_BT$ lower than other system's energy scales), $\alpha_{\text{Kelvin}}$ becomes a more precise formulation than the $\alpha_{\text{HM}}$ formula. The Kelvin and HM formulas may be equal under other conditions where the chemical potential $\mu \propto T$, but when $\mu \propto T^n$, they show an n-fold difference. This distinction is significant when considering metals at low temperatures [5]. Table 1 summarizes the above discussion.

The thermopower of a material is physically related to its specific heat capacity ($C_V$) and entropy ($S$). A high specific heat capacity indicates a large amount of heat stored in the material, while high entropy suggests a large number of available energy states for the electrons. Both factors influence the thermopower of a material. Despite the significance of thermopower, its direct dependence on specific heat capacity or entropy remains unresolved. Some studies have used specific heat capacity to calculate thermopower [6,7,8,9,10,11], while others have relied on entropy [12,13,14,15,16,17]. This study investigates the fundamental relationship between thermopower, specific heat capacity, and entropy, with the goal of resolving longstanding ambiguities in their interdependence.

Measuring entropy in mesoscopic systems (such as quantum dot systems and molecular junctions) is a fundamental challenge in condensed matter physics, as this thermodynamic quantity provides crucial insights into the accessible degrees of freedom in a system. Various methods have been proposed to extract entropy, but one of the most precise approaches is based on thermoelectric responses [18]. Since thermopower depends on entropy variations, it can be utilized as a tool to extract information about the entropy of mesoscopic systems. In this paper, we theoretically and experimentally demonstrate that thermopower is directly related to entropy.

We also determine which approach is more accurate and aim to identify conditions under which the approaches are equivalent and those under which one may lead to incorrect results. Using fundamental thermodynamic equations, we will assess the accuracy and applicability of each approach. This study assumes DC limit and static fields for the transport process, leading to a formula similar to Kelvin's.

We will assume a constant and energy-independent diffusion coefficient (see Section II. B). Our analysis shows that $\alpha_{\text{Kelvin}}$ can lead to inaccurate results even for the simple case of the thermopower of an electron gas [5]. Using the Onsager relation, we derive the Kelvin formula, highlighting the underlying assumptions often overlooked when using this method. Each method for calculating thermopower relies on specific assumptions, often treating parameters such as the electronic density of states ($\rho$), carrier velocity ($v$), or scattering relaxation time ($\tau$) as constants. These assumptions are not explicitly established at the outset of the methods. Essentially, different methods yield varying thermopower formulas based on conditions on $k$ and $\omega$. Comparing the final formulas derived from these methods reveals the underlying assumptions and their different outcomes. These parameters are crucial from both empirical and practical viewpoints, as experimentalists often use them to compare their results to theoretical values.

In the study of the thermodynamics and statistical properties of physical systems, internal energy is a fundamental thermodynamic quantity. According to the first and second laws of thermodynamics, the differential of the internal energy ($dU$) of an $N$-particle system can be expressed as:

$$dU = TdS - PdV + \mu dN, \quad (2)$$

where $S$ is entropy, $P$ is pressure, $V$ is volume and $N$ is number of particles in the system. As shown in Eq. (2), changes in internal energy depend on the extensive variables ($S, V, N$). An infinitesimal change in these variables results in a corresponding change in the internal energy of the system.

**Table 1. Different Thermopower Formulas:** In the Kubo formula, $\hat{J}_x^E(t - i\tilde{t})$ is the energy current operator depending on real time $t$ and the imaginary time $0 \leq \tilde{t} \leq \beta$, $\hat{J}_x$ is the charge current operator, $\mu(T)$ is the chemical potential, and $e$ is the charge of the particle (negative for electrons and positive for holes). In the Mott formula, $\sigma(\varepsilon)$ is electrical conductivity depending on energy of electrons, $n_e$ is the conduction band electron density for semiconductors, and $N_c = 2(mk_BT/2\pi\hbar^2)^{3/2}$ where $m$ is mass of carrier. In the high-frequency method, $\omega_c$ is the largest characteristic frequency, $\Phi^{xx}$ is the thermoelectric operator, $\tau^{xx}$ is the stress tensor, and $\langle\Phi^{xx}\rangle_0 \equiv \lim_{T\to 0}\langle\Phi^{xx}\rangle$. In the fourth column, the check/cross mark indicates whether thermopower is proportional to the specific heat capacity.

| Name | Symbol | Formula | Related to $C_V$ | Remarks |
|---|---|---|---|---|
| Kubo | $\alpha_{\text{Kubo}}$ | $\dfrac{1}{T}\dfrac{\int_0^\infty dt \int_0^\beta d\tilde{t}\, \langle \hat{J}_x^E(t-i\tilde{t})\hat{J}_x(0)\rangle}{\int_0^\infty dt \int_0^\beta d\tilde{t}\, \langle \hat{J}_x(t-i\tilde{t})\hat{J}_x(0)\rangle} - \dfrac{\mu(T)}{eT}$ | No | Fast DC limit of $\alpha(\mathbf{k},\omega)$ means first $\mathbf{k}\to 0$ and then $\omega\to 0$ [1]. The Kubo formula is precise within transport theory and is beneficial for non-interacting systems but challenging for strongly interacting electron systems [19]. Appropriate for dissipative systems [2]. |
| Mott | $\alpha_{\text{Mott}}$ | $\dfrac{\pi^2}{3}k_B T \left[\dfrac{\partial\ln(\sigma(\varepsilon))}{\partial\varepsilon}\right]\dfrac{k_B}{e};$ $\left(\dfrac{k_B}{e}\left[\dfrac{5}{2} - \ln\left(\dfrac{n_e}{N_c}\right)\right]\right)$ | No | Suitable for metals, insulators, and intrinsic semiconductors [3]. The Mott result derives from the general Kubo formula in the limit of weak scattering [2]. The expression in parentheses is for insulators or intrinsic semiconductors. |
| Heikes-Mott | $\alpha_{\text{HM}}$ | $\dfrac{(\mu(0)-\mu(T))}{eT}; \dfrac{\mu(T)}{T} = -\left(\dfrac{\partial S}{\partial N}\right)_{U,V}$ | Yes | High-temperature limit of thermopower [4]. Appropriate for narrowband systems [2] and strongly correlated materials, such as layered sodium cobaltate Na$_x$Co$_2$O$_4$, [5,19,20,21]. |
| Kelvin | $\alpha_{\text{Kelvin}}$ | $\dfrac{1}{e}\left(\dfrac{\partial S}{\partial N}\right)_{T,V}, -\dfrac{1}{e}\left(\dfrac{\partial \mu}{\partial T}\right)_{N,V}$ | Yes | Slow DC limit of $\alpha(\mathbf{k},\omega)$ where $\omega\to 0$ taken before $\mathbf{k}\to 0$ [4,5]. |
| High Frequency | $\alpha^*$ | $\lim_{\omega\gg\omega_c, \mathbf{k}\to 0} S(\mathbf{k},\omega) = \dfrac{\langle\Phi^{xx}\rangle - \langle\Phi^{xx}\rangle_0}{T\langle\tau^{xx}\rangle}$ | No | Superfast AC limit ($\omega\to\infty$) [1,2]. This formalism is accurate when scattering is less important than the density of states and correlations [2]. |

In a system of particles with charge e (negative for electrons and positive for holes), the internal energy also depends on the total charge. An infinitesimal increase in charge results in a corresponding increase in internal energy. To incorporate this into the internal energy, we introduce the electrochemical potential, denoted as $\bar{\mu}$ [22,23]:

$$\bar{\mu} = \mu + e\phi, \quad (3)$$

where the first term $\mu$ is the standard chemical potential and the second term accounts for the electrostatic energy stored in the system. Here $\phi$ denotes the electrostatic potential. The differential of the Helmholtz free energy is then modified as [13]:

$$dF = -SdT - PdV + \bar{\mu}dN. \quad (4)$$

By utilizing Maxwell's relation and exploiting that $dF$ is an exact differential, we obtain:

$$\frac{\partial^2 F}{\partial T \partial N} = \frac{\partial^2 F}{\partial N \partial T}. \quad (5)$$

From Eq. (4), we have $\left(\dfrac{\partial F}{\partial N}\right)_{T,V} = \bar{\mu}$, and $\left(\dfrac{\partial F}{\partial T}\right)_{N,V} = -S$. Thus, Eq. (5) can be transformed into:

$$\left(\frac{\partial \bar{\mu}}{\partial T}\right)_{N,V} = -\left(\frac{\partial S}{\partial N}\right)_{T,V}. \quad (6)$$

This equation plays a crucial role in deriving a general expression for the thermopower formula.

## II. THERMOPOWER FORMULAE

### A. Onsager to Kelvin

The Kelvin formula is often derived from thermodynamic relations where the underlying assumptions are not obvious. To clarify this matter, we will derive the thermopower coefficient using the Onsager formula:

$$\alpha = \frac{1}{T}\frac{K_1}{K_0}, \quad (7)$$

where:

$$K_n = e^{2-n} \int \frac{\rho(\varepsilon_k) d\varepsilon_k}{4\pi^3} \left(-\frac{\partial f}{\partial \varepsilon_k}\right) \tau(\varepsilon_k) \frac{1}{3} v_k^2 (\varepsilon_k - \mu)^n . \quad (8)$$

Here, $\varepsilon_k$ and $\rho(\varepsilon_k)$ represent the energy and density of states of the carriers, respectively, $v_k = |\boldsymbol{v_k}|$ denotes the magnitude of the band velocity, and $f$ is the Fermi-Dirac distribution function. Assuming that $v_k^2 \tau$ is energy-independent, at a given temperature $T$ and for a constant volume $V$, we can express $\alpha$ as:

$$\alpha = \frac{1}{eT} \frac{\frac{d}{d\mu} \int \varepsilon_k f(\varepsilon_k) \rho(\varepsilon_k) d\varepsilon_k}{\frac{d}{d\mu} \int f(\varepsilon_k) \rho(\varepsilon_k) d\varepsilon_k} - \frac{\mu}{eT}$$

$$= \frac{1}{eT} \left( \frac{\frac{dU}{d\mu} - \mu \frac{dN}{d\mu}}{\frac{dN}{d\mu}} \right)_{T,V}. \quad (9)$$

By employing Eq. (B7) in Appendix B, we obtain

$$\alpha = \frac{1}{eT} \frac{T \left(\frac{\partial S}{\partial \mu}\right)_{T,V}}{\left(\frac{\partial N}{\partial \mu}\right)_{T,V}} = \frac{1}{e} \left(\frac{\partial S}{\partial N}\right)_{T,V}. \quad (10)$$

This derivation leads us to the Kelvin formula from the Onsager formula and shows that the underlying assumption is the energy-independence of $v_k^2 \tau$, which is equivalent to assuming a constant diffusion coefficient $D$ in thermodynamic derivations (see Sec. II. B).

### B. Thermodynamic Derivation

Thermopower is typically defined as the ratio of the electric potential gradient to the temperature gradient across a material. However, in systems with charged particles, thermopower is redefined to account for the electrochemical potential gradient induced by a temperature gradient [24]:

$$\alpha = \frac{1}{|e|} \frac{\boldsymbol{\nabla}\bar{\mu}}{\boldsymbol{\nabla}T}, \quad (11)$$

where $|e| = 1.6 \times 10^{-19}$ C, applicable to both electrons and holes. To derive an expression for $\alpha$, we write the electric current density as a function of $\boldsymbol{\nabla}\bar{\mu}$ and $\boldsymbol{\nabla}T$. Consider a drift-diffusion model describing the flow of electric current through a material, which arises from both a temperature gradient applied across its ends and an external electric field. The total electric current density is the sum of three types of currents: diffusion current, drift current, and thermopower current induced by the temperature gradient. The total current density can be expressed as:

$$\boldsymbol{J} = -eD\boldsymbol{\nabla}n - \sigma\boldsymbol{\nabla}\phi + \alpha\sigma\boldsymbol{\nabla}T$$

$$= -eD\left(\frac{\partial n(T,\mu)}{\partial \mu}\boldsymbol{\nabla}\mu + \frac{\partial n(T,\mu)}{\partial T}\boldsymbol{\nabla}T\right)$$
$$-\sigma\boldsymbol{\nabla}\phi + \alpha\sigma\boldsymbol{\nabla}T. \quad (12)$$

Here, $D$ is the diffusion coefficient, $\sigma$ is the electrical conductivity of the material, $n$ is the charge carrier density (electrons or holes). In Eq. (12), the first term represents the diffusion current density, the second term represents the drift current density, and the last term arises from the voltage induced by the temperature gradient. The diffusion coefficient $D$ and the electrical conductivity σ are generally energy- and temperature-dependent. However, they are assumed to be constants in this equation, similar to the approach taken when deriving the Kelvin formula from Onsager's relations. In this context, $D = \frac{1}{3}\langle v \rangle \lambda$ (where $\langle \boldsymbol{v} \rangle$ and $\lambda$ denote the mean velocity and mean free path of the carriers, respectively.) is treated as a diffusion coefficient that is independent of temperature and energy.

In global equilibrium, $\boldsymbol{\nabla}\bar{\mu} = 0$ and consequently, $\boldsymbol{\nabla}\mu = -e\boldsymbol{\nabla}\phi$. The total current density can be simplified as:

$$\boldsymbol{J} = \left(e^2 D \frac{\partial n(T,\mu)}{\partial \mu} - \sigma\right)\boldsymbol{\nabla}\phi$$
$$+ \left(-eD\frac{\partial n(T,\mu)}{\partial T} + \alpha\sigma\right)\boldsymbol{\nabla}T. \quad (13)$$

Thermopower is often measured under open-circuit conditions, meaning the net current is zero. Given that the gradients $\boldsymbol{\nabla}\phi$ and $\boldsymbol{\nabla}T$ are external fields, they are independently applied to the system. To have $\boldsymbol{J} = 0$, both terms in Eq. (13) must be zero:

$$\sigma = e^2 D \frac{\partial n}{\partial \mu}, \quad (14)$$

$$\alpha = \frac{eD}{\sigma} \frac{\partial n}{\partial T}. \quad (15)$$

The first equation is the Einstein relation for the electrical conductivity of a material and the diffusion coefficient [23], and the second equation provides the thermopower of the materials. Utilizing Eqs. (6) and (14), we can obtain:

$$\left(\frac{\partial n}{\partial T}\right)_\mu = \left(\frac{\partial n}{\partial \mu}\right)_T \left(\frac{\partial \mu}{\partial T}\right)_n \left(\frac{\partial \bar{\mu}}{\partial \mu}\right) = \frac{\sigma}{e^2 D} \left(\frac{\partial s}{\partial n}\right)_T, \quad (16)$$

where we used $\partial\bar{\mu}/\partial\mu = 1$. Substituting this into Eq. (15), we achieve:

$$\alpha = \frac{1}{e}\left(\frac{\partial s(T,n)}{\partial n}\right)_T = \frac{1}{e}\left(\frac{\partial S(T,N)}{\partial N}\right)_{T,V}, \quad (17)$$

where, $N$ is the number of carriers, $n$ is the density of carriers, $s = S/V$ is entropy per volume. This is the main result of this section, demonstrating that thermopower is directly proportional to the partial derivative of the entropy density with respect to the particle density. Eq. (17) applies to charged particles. For neutral particles, such as magnons, the factor of $1/e$ is omitted, and thermopower is expressed in units of energy per temperature, such as eV/K.

As previously discussed, the drift-diffusion model we employed was independent of the energy of the particles under consideration, such as electrons, phonons, or photons. Although this might initially seem like a limitation, incorporating an energy-dependent diffusion coefficient $D(\varepsilon)$ typically results in only a minor prefactor of order unity [22]. Consequently, the drift-diffusion model remains an effective tool for qualitatively examining and analyzing electric and thermal transport phenomena. Even within the Sommerfeld expansion framework (energy-dependent model), the thermopower result is only slightly modified, reinforcing the utility of this approach. For example, for a Fermi gas, the thermopower calculated using the Kelvin formula (as detailed in Table 2 and Appendix A) is given by $\alpha_{\text{Kelvin}} = \frac{\pi^2}{6} \frac{k_B^2 T}{e \varepsilon_F}$. However, if the diffusion coefficient follows $D(\varepsilon) \propto \varepsilon$, applying the Sommerfeld expansion and using the Mott formula yields $\alpha_{\text{Mott}} = \frac{\pi^2}{3} \frac{k_B^2 T}{e \varepsilon_F}$, which is twice the value obtained from the Kelvin formula. Furthermore, if the density of states follows $\rho(\varepsilon) \propto \sqrt{\varepsilon}$, the thermopower increases to $\alpha = \frac{\pi^2}{2} \frac{k_B^2 T}{e \varepsilon_F}$.

Despite this limitation, our model offers several key advantages, including simplicity and broad applicability. It can readily be applied to magnetic semiconductors, where a combination of electrons and magnons can be modeled, allowing for the incorporation of the magnon-drag thermopower effect, as discussed in Section V. Additionally, the model can be extended to systems involving multiple particle species, such as electrons and holes in semiconductors, or spin-up and spin-down electrons in the fields of spintronics and spin-caloritronics, as well as in superconducting materials.

Moreover, the model's clarity regarding its underlying assumptions is particularly valuable for experimentalists. By explicitly defining the constant quantities involved, researchers can more easily assess the suitability of the model for their specific system based on the experimental parameters they control.

Entropy can be expressed in terms of specific heat capacity through the following thermodynamic relation [13]:

$$S = \int_0^T \frac{C_{V,N}}{T'} dT'. \quad (18)$$

Equation (18) illustrates that the entropy, and consequently the thermopower, at a given temperature $T$, is influenced by the specific heat across all temperatures below $T$. This suggests that the thermopower at a specific temperature depends on the accumulated thermal history of the system. This is analogous to the concept of the internal energy of a system, $U = \int_0^T C_V(T') dT'$, which also depends on the specific heat across all temperatures below $T$.

In some studies, thermopower is expressed directly in terms of specific heat per carriers as [6,8,11,25,26]:

$$\alpha = \frac{C_{V,N}(T)}{Ne}, \quad (19)$$

However, this relation is applicable only to systems where entropy and specific heat are identical. Mathematically, when specific heat has a power-law dependence on temperature as $C \propto T^r$, entropy and specific heat are proportional, with a proportionality factor of $r$. This behavior is observed, for example, in electrons in metals and degenerate semiconductors at low temperatures, which can be described by a Fermi gas or a Fermi liquid (with $C_e = \gamma T$). This is also observed for phonons at low temperatures, where $C_{ph} \propto T^d$, with $d$ representing the material's dimensionality.

Entropy, a thermodynamic property, indicates the degree of disorder within a system. At low temperatures, increasing the temperature leads to an increase in both specific heat and entropy. In this regime, the behaviors of entropy and specific heat are similar. However, as temperature increases further, systems with an infinite number of microstates exhibit a plateau in specific heat, while entropy continues to rise due to increasing disorder. This contrast is also evident in systems with a finite number of microstates (such as spin systems). In these systems, at high temperatures, specific heat decreases and approaches zero, while entropy continues to rise until it reaches a maximum value determined by the number of microstates, $(2\mathcal{S} + 1)^{\mathcal{N}}$, where $\mathcal{S}$ and $\mathcal{N}$ represent the spin quantum number and number of spins, respectively.

The key insight of our study is that thermopower, like entropy, is associated with the degree of disorder within a system. As disorder increases with temperature, thermopower generally increases as well.

For systems with a finite number of microstates, thermopower has an upper limit that is proportional to the logarithm of the number of microstates.

**Table 2. Specific heat per volume ($c_V = C_V/V$), entropy per volume ($s = S/V$), and thermopower for various systems:** In ideal gases, $\lambda_{th}$ is the thermal wavelength, and $n$ is representing particle density. In single-molecule junctions, $N(\varepsilon_0)$ is the occupation probability for an electron with energy $\varepsilon_0$ (Fermi-Dirac distribution function). The last column indicates that in Fermi gases and Fermi liquids, where specific heat exhibits power-law behavior, thermopower is proportional to specific heat. In a single-molecule junction, thermopower depends on the electronic specific heat ($c_V^e$) and the transmission coefficient. In other systems, it is proportional to entropy. The last row represents a spin system of MnTe with $\mathcal{N}$ number of spins, demonstrating the magnonic specific heat ($c_m$), entropy ($s_m$), and thermopower ($\alpha_m$) per volume. Detailed calculations are presented in the Appendix A.

| Systems | $c_V$ | $s$ | $\alpha$ | $\alpha \propto c_V$ or $s$ |
|---|---|---|---|---|
| Ideal gases [13] | $\frac{3}{2}nk_B$ | $nk_B\left[\frac{5}{2} - \ln(n\lambda_{th}^3)\right]$ | $k_B\left[\frac{3}{2} + \ln(n\lambda_{th}^3)\right]$ | $s$ |
| Fermi gas ($T \to 0$) [3,27] Fermi liquid[†] ($T \to \infty$) [3,11] | $\frac{mk_F}{3\hbar^2}k_B^2T = \frac{\pi^2}{2}\frac{nk_B^2T}{\varepsilon_F}$ | $\frac{mk_F}{3\hbar^2}k_B^2T = \frac{\pi^2}{2}\frac{nk_B^2T}{\varepsilon_F}$ | $\frac{\pi^2}{6}\frac{k_BT}{\varepsilon_F}\frac{k_B}{e}$ | $c_V, s$ |
| Single-molecule junction (Nanoscale) [28] | $\frac{\varepsilon_0^2}{k_BT^2V}N(\varepsilon_0)(1-N(\varepsilon_0))$ | $-\frac{k_B}{V}\left[\begin{array}{c}N(\varepsilon_0)\ln N(\varepsilon_0) \\ +(1-N(\varepsilon_0))\ln(1-N(\varepsilon_0))\end{array}\right]$ | $-\frac{k_B}{eV}[\ln N(\varepsilon_0) - \ln(1-N(\varepsilon_0))]$ $= \left(\frac{\varepsilon_0 - \mu}{k_BT}\right)\frac{k_B}{e}$ | $s$ |
| Magnetic spin system at high temperature ($T \to \infty$) [29] | $c_m \sim T^{-2}$ | $s_m \sim \mathcal{N}k_B \ln 6$ | $\alpha_m \sim k_B \ln 6$ | $s$ |

[†]We use $m^*$ for the Fermi liquid.

In Table 2, we represent a list of various systems and express their entropy and specific heat as functions of temperature. For systems where specific heat has a power-law dependence on temperature - by which we specifically mean a positive exponent of temperature - such as Fermi gases and Fermi liquids, entropy is proportional to specific heat. Therefore, in these cases, thermopower can be equivalently expressed using specific heat. However, for systems where specific heat does not follow a power-law temperature dependence - such as ideal gases, single-molecule junction, and magnetic spin systems - entropy and specific heat exhibit distinct behaviors, making thermopower more accurately represented by entropy.

At the nanoscale, interfaces can play a critical role in shaping thermopower behavior. For example, in single-molecule junctions - where a discrete energy level connects two leads - the transmission coefficient governs the rate at which charge carriers enter and exit the molecular state. This, in turn, determines the occupation probability and the effective chemical potential at a given temperature. As we show in the case study, the thermopower in such systems is primarily determined by the alignment of the molecular energy level with the chemical potential and is more accurately described by entropy rather than specific heat (see Section VII and Appendix E for details).

As we will discuss in the next section and conceptually demonstrate in [29], a more pronounced difference between the two quantities occurs in magnetic materials above the phase transition temperature.

### III. MAGNON THERMOPOWER

For magnetic systems with long-range magnetic orders, the spin excitations are magnon quasiparticles. In unfrustrated lattices, these quasiparticles are dispersive and propagate through the system with mobility and leading to a magnon conductivity. In general, for a magnetic system subject to a temperature gradient and a chemical potential gradient, the magnon current density is given by:

$$\boldsymbol{J}_m = D_m \boldsymbol{\nabla} n_m - \sigma_m \boldsymbol{\nabla}\mu_m - \alpha_m \sigma_m \boldsymbol{\nabla} T$$
$$= \left(D_m \frac{\partial n_m}{\partial \mu_m} - \sigma_m\right)\boldsymbol{\nabla}\mu_m + \left(D_m \frac{\partial n_m}{\partial T} - \alpha_m \sigma_m\right)\boldsymbol{\nabla} T, \quad (20)$$

where $n_m(T, \mu_m)$ is the magnon density, $D_m$ is the magnon diffusion coefficient, and $\mu_m$ is the magnon's chemical potential. The last term in the first line of Eq. (20) is the magnon current density is induced by a temperature gradient due to a thermopower effect. By setting $\boldsymbol{J}_m = 0$, we obtain:

$$\sigma_m = D_m \frac{\partial n_m}{\partial \mu_m}, \quad (21)$$

$$\alpha_m = \frac{D_m}{\sigma_m} \frac{\partial n_m}{\partial T}. \quad (22)$$

Employing the Gibbs-Duhem relation for magnons, i.e. $\frac{\partial n_m}{\partial T} = \frac{\partial s_m}{\partial \mu_m}$, where $s_m$ is the entropy density of the magnonic system, and utilizing $\frac{\partial s_m}{\partial \mu_m} = \frac{\partial s_m}{\partial n_m} \frac{\partial n_m}{\partial \mu_m}$ we reach to:

$$\alpha_m = \left(\frac{D_m}{\sigma_m} \frac{\partial n_m}{\partial \mu_m}\right) \frac{\partial s_m}{\partial n_m}. \quad (23)$$

According to the Einstein relation (21) for a magnonic system, the expression in the bracket (23) is equal to 1, and therefore the magnon thermopower reads:

$$\alpha_m = \frac{\partial s_m}{\partial n_m}. \quad (24)$$

Similar to the electron thermopower, the magnon thermopower is proportional to the changes in magnonic entropy to magnon density.

For a spin system with ferromagnetic order, at low temperatures, i.e., for $T \ll \mathcal{J}$ with $\mathcal{J}$ being the exchange interaction, magnons have parabolic dispersion, and the magnonic specific heat is given by the Bloch law, i.e. $c_m \sim T^{3/2}$ [13]. As we discussed in the previous section, here the specific heat has a power-law dependence on the temperature, and the entropy is identical to the specific heat. Therefore, at low temperatures, the magnon thermopower is proportional to the magnon specific heat [6,8]. According to the Bloch law, magnon density also has the same power-law dependence on temperature, i.e., $n_m \sim T^{3/2}$, which finally leads to a constant magnonic thermopower at low temperatures. By increasing the temperature, the Bloch law is violated and the magnon specific heat has a more complex temperature dependence. In this case, entropy controls the behavior of the magnonic thermopower.

At temperatures higher than the critical temperature, the magnetization of the system vanishes, indicating that the long-range order disappears and magnons are quenched by the temperature. In this region, the system is a paramagnet and although the magnons are removed, the system still has thermopower. This is because of the presence of short-range orders above the critical point. There may be different scenarios to quantify the thermopower of magnetic systems at temperatures higher than the critical temperature [11,30,31,32]. One considers paramagnons as collective excitations of short-range orders and considers their contributions to the entropy. Although in presenting the thermopower formula, we start with a magnon current density, Eq. (24) is general and gives the thermopower at all temperatures below and above the critical point, with $s_m$ being the entropy of the interacting spin-lattice and $n_m$ serving as the number of spins per volume.

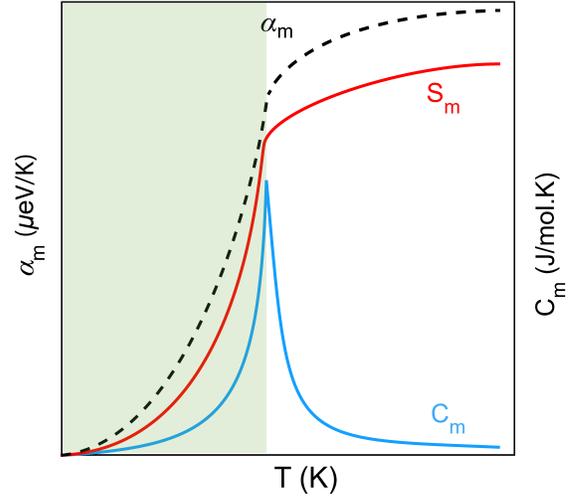

Figure 1: Schematic representation of the specific heat, entropy, and thermopower of a magnetic semiconductor as a function of temperature. At the critical temperature, a phase transition occurs from the ferromagnetic (or antiferromagnetic) to the paramagnetic phase. Above the phase transition temperature, the specific heat declines, while the entropy, which depends on the integral of $C_m/T$ with respect to temperature, does not decline. Consequently, thermopower, which depends on entropy, also does not decline. Thus, in such cases, the thermopower cannot be explained by specific heat, and entropy provides a more accurate explanation of its trend [11,29,32].

Figure 1 illustrates the conceptual behavior of magnon thermopower in a magnetic semiconductor. It shows that thermopower, specific heat, and entropy follow similar trends below the magnetic phase transition. However, above the phase transition temperature, while the magnon specific heat vanishes, the entropy and thus the thermopower do not decline and continue to follow similar trends. This behavior has been experimentally observed in several magnetic materials, such as MnTe and MnSe [33,34], and layered oxides $Na_xCo_2O_4$ [19].

In the last row of Table 2, we discuss the thermodynamic properties of a magnetic system such as MnTe, which is recognized as a promising thermoelectric material at high temperatures. The thermopower of MnTe includes an additional term associated with magnons that propagate through the Mn spin-lattice and interact with the charge carriers. This interaction leads to a magnon-hole drag phenomenon, resulting in an enhancement of the thermopower near the Néel temperature, where the system undergoes a phase transition from an ordered antiferromagnetic phase to a paramagnetic disordered phase. It has been shown that the drag thermopower in MnTe is directly proportional to the magnonic specific heat per carrier, $C_m$ [6]:

$$\alpha_{md} = \frac{C_m \langle R \rangle}{3e}, \qquad (25)$$

where $\langle R \rangle$ represents the average ratio of the momentum transferred from the magnons to carriers. Since MnTe is a p-type semiconductor with hole carriers, $e > 0$ in Eq. (25), resulting in a positive magnon thermopower.

It has been demonstrated that at low temperatures, where magnons are considered as an ideal Bose gas with a parabolic spectrum, the magnon thermopower is given by the following equation [8,11,26]:

$$\alpha_{md} = \frac{2}{3} \frac{C_m}{n_e e} \frac{\tau_m}{\tau_m + \tau_{me}}, \qquad (26)$$

where, $\tau_m$ is the magnon relaxation time due to scattering by everything except charge carriers, and $\tau_{me}$ is the magnon relaxation time due to scattering only by charge carriers.

At high temperatures above the Néel temperature, the specific heat of the magnetic spin system exhibits a power-law decay, proportional to $\sim T^{-2}$ [29]. According to Eq. (25), above the transition temperature, the drag thermopower should decrease with increasing temperature and approach zero at high temperatures. However, this prediction contrasts with the experimental observations (see Figure 2, square symbols), where the drag thermopower extends beyond the Néel temperature without a decline.

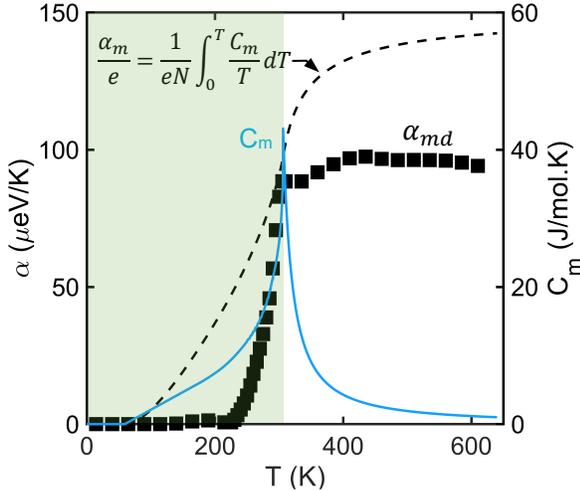

Figure 2: Magnon thermopower, specific heat, and magnon-drag thermopower of a magnetic semiconductor (MnTe) as a function of temperature. A phase transition occurs from the antiferromagnetic ordered phase (colored region) to the paramagnetic disordered phase at the Néel temperature. Below the Néel temperature, the magnon and magnon-drag thermopower align with the trend of the entropy ($\int_0^T C_m/T' dT'$). Above the Néel temperature, paramagnon and paramagnon-drag thermopower follow a similar trend [11,29, 32].

In the following section, we demonstrate that according to our analysis, the drag thermopower is proportional to the entropy of the magnetic spin-lattice. As the temperature approaches infinity, entropy converges to a constant value given by $\mathcal{N} k_B \ln(2S + 1)$.

For example, the magnetic thermopower for the MnTe system with $S = 5/2$ is given by $k_B \ln 6$ or approximately ~154 µV/K, which falls within the expected range based on experimental values of the magnon-drag thermopower [11,29,35].

### IV. MAGNON-DRAG THERMOPOWER

In general, scattering can be described by a simple equation: $d\mathbf{v}/dt = -\mathbf{v}/\tau$, where $\mathbf{v}$ is the drift velocity of particles [36]. The phenomenological equation for the momentum of the fluid can be written as:

$$\frac{d\mathbf{p}_e}{dt} = n_e e (\mathbf{E} - \alpha \nabla T) - \frac{\mathbf{p}_e}{\tau_e} - \frac{\mathbf{p}_e}{\tau_{em}} + \frac{\mathbf{p}_m}{\tau_{me}}, \qquad (27)$$

$$\frac{d\mathbf{p}_m}{dt} = -n_m \alpha_m \nabla T - \frac{\mathbf{p}_m}{\tau_m} - \frac{\mathbf{p}_m}{\tau_{me}} + \frac{\mathbf{p}_e}{\tau_{em}}, \qquad (28)$$

where $\mathbf{E}$ is the electric field, $\mathbf{p}_e = m\mathbf{v}_e$ is the momentum of the electron with carrier effective mass $m$ and drift velocity $\mathbf{v}_e$, $\mathbf{p}_m$ is the magnon momentum, $\alpha$ is the electronic diffusion thermopower, $\alpha_m$ is the magnon thermopower, and $\tau_e$ is the electron momentum relaxation time due to all scattering mechanisms except scattering by magnons. $\tau_m$, as defined earlier, is the total momentum relaxation time for magnons due to all scattering mechanisms except scattering by electrons. $\tau_{me}$, also defined earlier, is the magnon relaxation time due to scattering by electrons. $\tau_{em}$ is the electron relaxation time due to scattering by magnons. In this two-fluid model the electron fluid (treated here non-relativistically) and the magnon fluid (treated non-relativistically and relativistically) interact through mutual drag terms. We define the momentum transfer times $\tau_{em}$ and $\tau_{me}$ which characterize the exchange of momentum between electrons and magnons.

To provide a comprehensive description of magnon-drag thermopower across different magnetic systems, we consider both massive and massless magnons in our theory. This distinction allows the model to capture a wide range of physical regimes, including the low-temperature limit in antiferromagnets (AFMs), where magnons exhibit a linear dispersion relation and behave as massless relativistic quasiparticles. In contrast, ferromagnets (FMs) are characterized by a quadratic magnon dispersion, which enables a massive treatment of magnons. By developing separate formulations for each case, our theory remains general and applicable at all temperatures while ensuring consistency with known low-temperature magnon behavior.

## A. MASSIVE MAGNON
## (NEWTONIAN METHOD)

In the hydrodynamic theory, magnons and electrons are modeled as two interpenetrating fluids [26], and Galilean invariance is assumed such that the electrons and magnons are described by a single parabolic band. Furthermore, Umklapp and magnon non-conserving processes are neglected. The momentum of a magnon with group velocity $\boldsymbol{v}_m$ and mass $M$ (which characterizes the response of magnons to external forces based on their dispersion relation), written as $\boldsymbol{p}_m = M\boldsymbol{v}_m$. From Eqs. (27) and (28) under steady-state conditions $(d\boldsymbol{v}_m/dt = 0, d\boldsymbol{v}_e/dt = 0)$ and for zero electric current ($\boldsymbol{v}_e = 0$), we get:

$$\left[\frac{n_m \alpha_m}{M}\left(\frac{1}{\tau_m}+\frac{1}{\tau_{me}}\right)^{-1} + \frac{n_e e}{M}\tau_{me}\alpha\right]\boldsymbol{\nabla}T = \frac{n_e e}{M}\tau_{me}\boldsymbol{E}. \quad (29)$$

We use Eq. (1) to determine the total thermopower formula:

$$\frac{|\boldsymbol{E}|}{|\boldsymbol{\nabla}T|} = \frac{\alpha_m}{e}\left(\frac{n_m}{n_e}\right)\left(\frac{\tau_m}{\tau_m+\tau_{me}}\right) + \alpha. \quad (30)$$

So, the magnon-drag thermopower can be written as:

$$\alpha_{md} = \frac{\alpha_m}{e}\left(\frac{n_m}{n_e}\right)\left(\frac{\tau_m}{\tau_m+\tau_{me}}\right). \quad (31)$$

According to (24), we observe that magnon thermopower is related to the derivative of entropy with respect to the number of magnons. This correspondence for diffusion thermopower is a good approximation, especially for metals with quasi-particles and heavy quasi-holes at low temperatures [22]. Thus, we can express the thermopower for the magnon system similar to that of the electron system:

$$\alpha_{md} = \frac{1}{e}\frac{\partial s_m}{\partial n_m}\frac{n_m}{n_e}\left(\frac{\tau_m}{\tau_m+\tau_{me}}\right)$$
$$\sim \frac{1}{e}\frac{s_m}{n_e}\left(\frac{\tau_m}{\tau_m+\tau_{me}}\right). \quad (32)$$

One can show that, $\tau_{me} = \frac{k_{\rm B}T}{mv_m^2}\tau_{em}$ [6,37]. This relation is smaller by a factor of 2, because Zanmarchi and Haas [37] assumed a single magnon mode. The magnon modes in AFMs are usually doubly degenerate, resulting in $\tau_{me}$ two times larger [11]. Using this relation, we can rewrite the first equality in Eq. (32) for AFMs as follows:

$$\alpha_{md} = \frac{1}{e}\frac{\partial s_m}{\partial n_m}\frac{n_m}{n_e}\left(\frac{mv_m^2}{k_{\rm B}T}\frac{\tau_m}{\tau_{em}}\right)\left(\frac{1}{1+\tau_m/\tau_{me}}\right). \quad (33)$$

The second parenthesis, typically minor, represents a second-order correction resulting from the momentum transferred from magnons to electrons before being randomized. Therefore, as indicated by Eq. (32), magnon drag thermopower is influenced by magnon entropy per carrier rather than by magnon specific heat. Our findings change those presented by Sugihara in Ref. [6] and fundamentally revising the prior understanding of magnon-drag thermopower.

## B. MASSLESS MAGNON
## (RELATIVISTIC METHOD)

In ferromagnetic materials, magnon-drag thermopower is often described using semiclassical hydrodynamic models as described in the previous section, where magnons and electrons are treated as two interpenetrating Newtonian fluids. These models rely on assumptions such as a parabolic magnon dispersion $\omega_q = \mathcal{D}q^2$, the existence of an effective magnon mass $M$, and classical momentum relations $\boldsymbol{F} = m\boldsymbol{a}$ and $\boldsymbol{p} = m\boldsymbol{v}$. While these assumptions are reasonable in ferromagnets at low temperatures (where long-wavelength magnons dominate and the quadratic dispersion is valid), they break down in AFM systems, particularly at low energy. In AFMs, magnons exhibit a linear dispersion $\omega_q = \hbar v_m q$, where $v_m$ is the spin excitation (magnon) group velocity. This behavior classifies them as massless, relativistic bosonic quasiparticles, rendering Newtonian formulations physically inconsistent due to the absence of a meaningful magnon mass and the breakdown of the $\boldsymbol{p} = m\boldsymbol{v}$ relation.

To properly describe the thermoelectric response in AFMs, we adopt a relativistic fluid approach, modeling the magnon system using the energy-momentum tensor of a perfect relativistic fluid. This formulation inherently accommodates the massless nature of AFM magnons and provides a consistent framework for capturing momentum exchange with the electronic system. The dynamics of the magnon fluid are governed by the energy-momentum conservation law:

$$\partial_t T^{\iota\nu} = F^\nu, \quad (34)$$

where $T^{\iota\nu}$ is the energy-momentum tensor and $F^\nu$ accounts for external perturbations such as temperature gradients and scattering-induced momentum transfer. For a perfect fluid of massless bosons, the energy-momentum tensor is given by:

$$T^{\iota\nu} = (\epsilon + P)u^\iota u^\nu + P\eta^{\iota\nu}, \quad (35)$$

where $\epsilon$ is the energy density of the magnon fluid, $P$ is the pressure, $u^\iota$ is the four-velocity of the fluid, and $\eta^{\iota\nu}$ is the Minkowski metric [38]. In systems with linear dispersion, the pressure is related to the energy density via the relation $P = \epsilon/3$. The momentum density of the magnon fluid is given by the mixed time-space component of the energy-momentum tensor:

$$\boldsymbol{p}_m = T^{0i} = (\epsilon + P)\boldsymbol{v}_m$$

We now construct the coupled momentum balance equations for electrons and magnons under a set of physically reasonable assumptions. First, we assume the system is near equilibrium and spatially uniform, allowing us to neglect spatial derivatives of thermodynamic quantities beyond the applied gradient. The magnon dispersion is taken to be linear, $\omega_q = \hbar v_m q$, which is characteristic of AFM magnons at low energies. The electron and magnon fluids interact via momentum exchange processes characterized by relaxation times $\tau_{em}$ and $\tau_{me}$, corresponding to electron-to-magnon and magnon-to-electron scattering, respectively. Electrons experience both an electric field $\boldsymbol{E}$ and a temperature gradient $\boldsymbol{\nabla}T$, and are characterized by an intrinsic thermopower $\alpha$. In contrast, magnons, being charge-neutral, respond only to the thermal gradient and possess a thermopower $\alpha_m$. We impose an open-circuit boundary condition, such that there is no net electron drift ($\boldsymbol{v}_e = 0$). Finally, we consider the steady-state regime, where the time derivatives of both the electron and magnon momentum densities vanish: $d\boldsymbol{p}_e/dt = d\boldsymbol{p}_m/dt = 0$. These assumptions allow us to derive a consistent and closed-form solution for the magnon-drag thermopower in systems with linearly dispersing magnons.

The overall two-fluid coupling scenario is illustrated schematically in Figure 3, which provides a conceptual guide for the derivation that follows.

Under these assumptions, similar to Eqs. (27) and (28), the relativistic momentum equations for electrons and magnons become:

$$\frac{d\boldsymbol{p}_e}{dt} = n_e e(\boldsymbol{E} - \alpha\boldsymbol{\nabla}T) - \frac{\boldsymbol{p}_e}{\tau_e} - \frac{\boldsymbol{p}_e}{\tau_{em}} + (\epsilon + P)\frac{\boldsymbol{v}_m}{\tau_{me}}, \quad (36)$$

$$\frac{d\boldsymbol{p}_m}{dt} = -n_m \alpha_m \boldsymbol{\nabla}T - (\epsilon + P)\left(\frac{\boldsymbol{v}_m}{\tau_m} + \frac{\boldsymbol{v}_m}{\tau_{me}}\right) + \frac{\boldsymbol{p}_e}{\tau_{em}}, \quad (37)$$

In the open-circuit and steady-state limit ($\boldsymbol{v}_e = 0$, $d\boldsymbol{p}/dt = 0$), the equations simplify to:

$$-n_e e(\boldsymbol{E} - \alpha\boldsymbol{\nabla}T)\tau_{me} = (\epsilon + P)\boldsymbol{v}_m \quad (38)$$

$$n_m \alpha_m \boldsymbol{\nabla}T = (\epsilon + P)\left(\frac{\boldsymbol{v}_m}{\tau_m} + \frac{\boldsymbol{v}_m}{\tau_{me}}\right), \quad (39)$$

Solving Eq. (39) for $\boldsymbol{v}_m$, we find:

$$\boldsymbol{v}_m = \frac{n_m \alpha_m}{(\epsilon + P)}\boldsymbol{\nabla}T\left(\frac{1}{\tau_m} + \frac{1}{\tau_{me}}\right)^{-1} \quad (40)$$

Substituting into Eq. (38) and solving for the induced electric field, we obtain:

$$\frac{|\boldsymbol{E}|}{|\boldsymbol{\nabla}T|} = \alpha + \frac{n_m \alpha_m}{n_e e} \cdot \frac{1}{\tau_{me}}\left(\frac{1}{\tau_m} + \frac{1}{\tau_{me}}\right)^{-1}. \quad (41)$$

The resulting magnon-drag thermopower is thus:

$$\alpha_{md} = \frac{\alpha_m}{e}\left(\frac{n_m}{n_e}\right)\left(\frac{\tau_m}{\tau_m + \tau_{me}}\right) \quad (42)$$

This expression matches the form previously derived in Newtonian fluid models for ferromagnets, but here it emerges from a relativistically consistent framework that fully accounts for the massless, linearly dispersing nature of antiferromagnetic magnons. It confirms that the commonly used entropy-based expression for magnon-drag thermopower remains valid, provided it is interpreted through a more general theoretical lens. As illustrated in Figure 3, magnons driven by a thermal gradient transfer momentum to the electron fluid and induce an electric field, even under open-circuit conditions. This entropy-mediated momentum exchange forms the physical basis of magnon-drag thermopower in both ferromagnetic and antiferromagnetic systems.

Moreover, our result clarifies the limitations of traditional models that rely on effective magnon mass or the substitution $\boldsymbol{p} = \hbar \boldsymbol{q} \sim M\boldsymbol{v}$. While such models may appear to work phenomenologically, they are conceptually flawed for AFMs, where group velocity is dispersion-independent and the quasiparticles do not accelerate under external forces in the Newtonian sense. By deriving the magnon-drag thermopower from relativistic conservation laws, we establish a general and robust foundation for interpreting experimental results in both AFM and FM systems, across a wide temperature range.

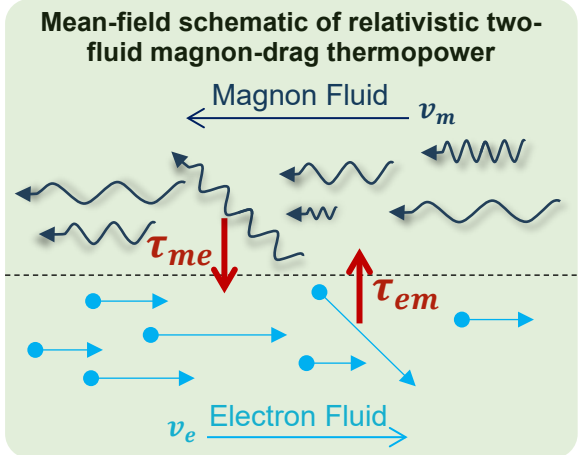

Figure 3: This diagram illustrates how, in a mean-field sense, the electron and magnon fluids exchange momentum through effective interactions. A thermal gradient $\nabla T$ drives a magnon flow $\boldsymbol{v}_m$, which transfers momentum to the electron fluid and drags back on the electrons, inducing an electric field $\boldsymbol{E}$, even when magnons are massless and governed by relativistic dynamics. The electron and magnon fluids interact via scattering processes characterized by relaxation times $\tau_{me}$ and $\tau_{em}$. $\tau_{me}$ quantifies how quickly magnons lose momentum to electrons, while $\tau_{em}$ describes how quickly electrons lose momentum to magnons. In this schematic, straight arrows of varying lengths represent electrons with different Fermi wavelengths (~0.5–1 nm), while wavy arrows with differing wavelengths represent magnons. The thermal wavelength of magnons typically ranges from ~10 nm to several hundred nanometers, depending on the dispersion: it scales as $T^{-1}$ for relativistic (linear) magnons in antiferromagnets and as $T^{-1/2}$ for non-relativistic (parabolic) magnons in ferromagnets, reflecting their collective, long-wavelength nature at low temperatures. In drag-dominated regimes, magnons are driven by $\nabla T$, while electrons respond to both $\nabla T$ and $\boldsymbol{E}$, and under open-circuit conditions ($\boldsymbol{v}_e = 0$), the back-action from the magnon flow leads to the development of thermoelectric voltage. Depending on the relative signs of $\nabla T$, $\boldsymbol{E}$, and the magnon thermopower $\alpha_m$, magnons may flow in the same or opposite direction as electrons. The resulting electric field typically opposes the thermal gradient, as observed in thermopower measurements.

As mentioned in [6] and [26], at low temperatures, the magnon drag thermopower is related to the specific heat (see Eqs. (25) and (26)). It has been previously established that the specific heat behaves as $C_m \sim T^{3/2}$ at low temperatures. Since $S_m$ is the integral of $C_m$ with respect to $T$, entropy is proportional to specific heat. Consequently, according to Eq. (32), which we derived for magnon drag, the magnon drag thermopower at low temperature is also proportional to the specific heat (see Figure 2). However, at high temperatures, the situation differs, and $C_m$ exhibits different behavior ($T \to \infty$, $C_m \sim T^{-2}$). Therefore, the above-mentioned approximation cannot be applied, and the behavior of $C_m$ with respect to $T$ does not exhibit a similar pattern to that of $S_m$. In the FM or AFM phase, $S_m$ is linked to magnons, whereas in the paramagnetic phase, it is associated with paramagnons. Both magnon and paramagnon drag thermopowers are determined by the derivative of entropy with respect to the number of magnons or paramagnons.

## V. CASE STUDY 1: THERMODYNAMIC AND THERMOPOWER OF MAGNETIC MATERIALS

In Table 3, we report various properties for several magnetic materials (Fe, Co, Ni, CrSb, MnSb, MnTe) and compare them. The table includes values for the Heisenberg exchange interaction ($\mathcal{J}$), carrier concentration $n_e$, spin stiffness $\mathcal{D}$, spin excitation (magnon) velocity $v_m$, magnon frequency ($\omega = \mathcal{E}_q/\hbar$), spin quantum number ($\mathcal{S}$), magnetic ion, spin-diffusion constant $\Lambda$, magnon relaxation time $\tau_m$, maximum experimental magnon specific heat ($C_m$), magnon entropy ($S_m$), magnon thermopower ($\alpha_m$), experimental magnon-electron drag thermopower ($\alpha_{md}$), transition temperature ($T_C$ or $T_N$), lattice structure, lattice parameters, magnetic ordering, and material type.

According to Eq. (33), the drag thermopower depends on several factors, including the magnon group velocity ($v_m$), temperature, the ratio $\tau_m/\tau_{em}$, and magnonic entropy. The magnon group velocity is primarily influenced by the exchange interaction ($\mathcal{J}$) between magnetic ions, approximately given by $v_m \propto \mathcal{J} a \mathcal{S}/\hbar$ in AFMs which is independent of the magnon's wavevector, and $v_m \propto \mathcal{J} a^2 \mathcal{S} q/\hbar$ in FMs, where $q$ is the magnon's wavevector magnitude. Thus, a larger $\mathcal{J}$ value enhances the magnon group velocity, which in turn favors higher thermopower.

The average energy-independent magnon relaxation time ($\tau_m$), is calculated using the following relation:

$$\frac{1}{\tau_m(T)} = \frac{\int_0^\infty \left(1/\tau_m(\varepsilon_q)\right) N(\varepsilon_q) g(\varepsilon_q) d\varepsilon_q}{\int_0^\infty N(\varepsilon_q) g(\varepsilon_q) d\varepsilon_q}. \quad (43)$$

In this equation, $g(\varepsilon_q)$ is the magnon density of states, $N(\varepsilon_q)$ is the Bose-Einstein distribution function, and $\tau_m(\varepsilon_q)$ is the energy-dependent magnon relaxation time. This formulation integrates over all energy states to yield an effective relaxation time for magnons, accounting for contributions from different states weighted by their densities and distributions.

The energy-dependent relaxation time, $\tau_m(\varepsilon_q)$ or $\tau_m(q)$, ignoring Rayleigh scattering, is related to the parameter $\Lambda$ [39]:

$$\frac{1}{\tau_m(q)} = \Lambda q^2, \quad (44)$$

where $\Lambda$ is estimated using the following expression:

$$\Lambda = \sum_{i=1}^{Z} \mathcal{J}_i \mathcal{R}_i^2 \sqrt{S(S+1)}/\hbar. \quad (45)$$

Here, $\mathcal{J}_i$ is the exchange interaction between a magnetic atom and its $i^{th}$ neighbor, $Z$ is the number of neighbors considered, starting from the nearest and extending to the furthest, $\mathcal{R}_i$ is the distance of a magnetic atom from its $i^{th}$ neighbor, and $S$ is the spin quantum number.

While the spin diffusion constant ($\Lambda$) and the magnon diffusion coefficient ($D_m$) share the same units, they describe different physical processes. $\Lambda$ quantifies the spread of spin polarization in a material over time, which is relevant to spin transport phenomena. In contrast, $D_m$ characterizes how magnons, as collective excitations of spins, propagate and disperse within a material.

The magnon-drag thermopower is directly proportional to $\tau_m$. A longer $\tau_m$ generally results in higher $\alpha_{md}$, a trend exemplified by MnTe, which shows the highest magnon-drag thermopower among the materials studied, as shown in Table 3. According to Eq. (43), $\tau_m$ depends on parameters such as $\Lambda$, $\mathcal{D}$, $\omega$, and $v_m$. Calculations show that MnTe has the largest magnon relaxation time, which allows a greater portion of magnon entropy to transfer to electrons [11,25,34].

Magnon lifetimes are particularly long near the Néel temperature in MnTe, due to weaker magnon-magnon and magnon-phonon scattering. These long lifetimes allow MnTe to sustain significant drag thermopower even at higher temperatures, in contrast to many magnetic materials, where $\alpha_{md}$ drops sharply immediately above the transition temperature. This behavior is explained by the presence of long-lived paramagnons above $T_N$, which play a crucial role in sustaining thermopower behavior. In other materials, the relaxation time is often too short for paramagnons to effectively contribute to the drag effect, limiting $\alpha_{md}$. Thus, the high $\tau_m/\tau_{em}$ ratio in MnTe, particularly above $T_N$, is a key factor contributing to its relatively large thermopower [11,26,34].

**Table 3. Comparison of key properties for various magnetic materials, including Fe, Co, Ni, CrSb, MnSb, and MnTe:** Reported properties include the Heisenberg exchange interaction ($\mathcal{J}/k_B$ (Kelvin)) up to the first 10 shells for Fe, Co, and Ni; for CrSb, MnSb, and MnTe, values specific to their structures, the number of neighboring atoms for each exchange interaction (Z), carrier concentration ($n_e$), spin stiffness ($\mathcal{D}$), spin excitation velocity ($v_m$), frequency ($\omega = \mathcal{E}_q/\hbar$), spin quantum number ($S$), magnetic ion, spin-diffusion constant ($\Lambda$), magnon relaxation time ($\tau_m$), maximum experimental magnon specific heat ($C_m$), magnon entropy ($S_m$), magnon thermopower ($\alpha_m$), experimental magnon-electron thermopower data ($\alpha_{md}$), transition temperature ($T_C$ or $T_N$), lattice structure, lattice parameters, magnetic ordering, and material type.

| Materials | Fe | Co | Ni | CrSb | MnSb | MnTe |
|---|---|---|---|---|---|---|
| $\mathcal{J}_1/k_B$, (Z) | 225.91, (8) | 171.17, (12) | 32.50 (12) | −134.6, (2) | $\mathcal{J}_1$ | 21.5, (2) |
| $\mathcal{J}_2/k_B$, (Z) | 128.57, (6) | 1.74, (6) | 0.95 (6) | 22.85, (6) | $14.1 - \mathcal{J}_1/3$ | −0.67, (6) |
| $\mathcal{J}_3/k_B$, (Z) | −2.52, (12) | 18.30, (24) | 4.10, (24) | | $8.5 - \mathcal{J}_1/6$ | 2.87, (12) |
| $\mathcal{J}_4/k_B$, (Z) | −19.88, (24) | −14.20, (12) | 1.89, (12) | | 13.78 | |
| $\mathcal{J}_5/k_B$, (Z) | −23.03, (8) | 4.10, (24) | 0.47, (24) | | or | |
| $\mathcal{J}_6/k_B$, (Z) | 9.78, (6) | 6.78, (8) | −0.47, (8) | | $4S(\mathcal{J}_1 + 6\mathcal{J}_3) = 406$ | |
| $\mathcal{J}_7/k_B$, (Z) | 0.16, (24) | −3.79, (48) | 1.10, (48) | | $12S(\mathcal{J}_2 + 2\mathcal{J}_3) = 754$ | |
| $\mathcal{J}_8/k_B$, (Z) | 2.37, (24) | 1.89, (6) | −0.16, (6) | | $4S\mathcal{J}_4 = 110.2$ | |
| $\mathcal{J}_9/k_B$, (Z) | −5.05, (24) | 4.10, (12) | −1.74, (12) | | | |
| $\mathcal{J}_{10}/k_B$, (Z) | 29.50, (8) | 0.95, (24) | 0.16, (24) | | | |

| Refs in which $\mathcal{J}$'s are reported | [40] | [40] | [40] | [41] | [42] | [43] |
|---|---|---|---|---|---|---|
| $n_e$ (cm$^{-3}$) | $1.7 \times 10^{23}$ [44] | $1.81 \times 10^{23}$ [44] | $1.83 \times 10^{23}$ [44] | $7 \times 10^{21}$ [34] | $8 \times 10^{21}$ [34] | $1 \times 10^{19}$ [35] |
| $\mathcal{D}$ (FM) / $v_m$ (AFM) | 27 meVÅ$^2$ [26] | 43 meVÅ$^2$ [26] | 59 meVÅ$^2$ [26] | $1 \times 10^4$ m/s [34] | 5.55 meVÅ$^2$ [42] | $1.4 \times 10^4$ m/s [37] |
| $v_m$ (m/s) | $54\,q$ | $86\,q$ | $118\,q$ | $1 \times 10^4$ | $11.1\,q$ | $1.4 \times 10^4$ |
| $\omega$ at low temperatures | $\mathcal{D}q^2$ | $\mathcal{D}q^2$ | $\mathcal{D}q^2$ | $v_m q$ | $\mathcal{D}q^2$ | $v_m q$ |
| Spin number* $\mathcal{S}$ | 2 [45] | 1.5 [45] | 1 [45] | 1.5 [41] | 2 [45] | 2.5 [45] |
| Magnetic ion | Fe$^{2+}$ | Co$^{2+}$ | Ni$^{2+}$ | Cr$^{3+}$ | Mn$^{3+}$ | Mn$^{2+}$ |
| $\Lambda \times 10^{-7}$ (m$^2$/s) | 1827 | 907 | 140 | 263 | 161 | 60 |
| $\tau_m$ (fs) at $T_{C/N}$ | 0.021 | 0.051 | 1.01 | 0.042 | 0.086 | 1.94 |
| $C_m$ (J/mol.K) at $T_{C/N}$ | 50.4 | 19.6 | 10.1 | 73.6 | 22.4 | 42.4 |
| $S_m$ (J/mol.K) | 8.3 | 6.6 | 2.6 | 7.87 | 7.58 | 12.72 |
| $C_m/n_e$ (µeV/[mol.K.m$^{-3}$]) | $1.8 \times 10^{-3}$ | $6.7 \times 10^{-4}$ | $3.4 \times 10^{-4}$ | $6.6 \times 10^{-2}$ | $1.7 \times 10^{-2}$ | 26.53 |
| $S_m/n_e$ (µeV/[mol.K.m$^{-3}$]) | $3.13 \times 10^{-4}$ | $2.28 \times 10^{-4}$ | $8.98 \times 10^{-5}$ | $7.05 \times 10^{-3}$ | $5.94 \times 10^{-3}$ | 7.93 |
| $\alpha_m \sim S_m/n_m$ (µeV/K) | 86.0 | 68.5 | 27.5 | 81.7 | 78.6 | 132.0 |
| $\alpha_{md}$ (µV/K) | 30 at 300 K [46] | −40 at 1000 K [26] | −30 [26] | −27 [34] | −6 [34] | 100 [34] |
| Transition temperature (K) | $T_C = 1043$ [47] | $T_C = 1388$ [48] | $T_C = 633$ [49] | $T_N = 713$ [34] | $T_C = 587$ [51] | $T_N = 307$ [11] |
| Structure | BCC [50] | FCC [50] | FCC [50] | NiAs [41] | NiAs [42] | NiAs [11] |
| Lattice parameter (Å) | $a = 2.867$ [50] | $a = 3.429$ [50] | $a = 3.436$ [50] | $a = 4.12\ c = 5.46$ [51] | $a = 4.13\ c = 5.79$ [42] | $a = 4.15\ c = 6.71$ [52] |
| Magnetic ordering | FM [50] | FM [50] | FM [50] | AFM [34] | FM [34] | AFM [11] |
| Material type | Metal [50] | Metal [50] | Metal [50] | Semiconductor [34] | Semimetal [53] | Semiconductor [11] |

*The exact value of $\mathcal{S}$ may differ from the listed numbers. Without considering temperature effects, the spin state of the system is determined by the competition between pairing energy and the crystal field. However, when temperature effects are included, the spin state may change to minimize the system's free energy.

This higher magnon-drag thermopower also stems from the strong interaction between charge carriers and magnons, which leads to a very short $\tau_{em}$ near $T_N$ [33]. A small $\tau_{em}$ indicates a high electron momentum relaxation rate due to Umklapp scattering with magnons, enabling more efficient electron drag by magnons. In magnetic metals, this effect is often diminished due to momentum loss from electron-magnon Umklapp processes. Therefore, the large ratio of $\tau_m/\tau_{em}$ in MnTe is in agreement with its exceptionally high magnon-drag thermopower.

Comparing the properties of magnetic materials in Table 3 reveals that a larger exchange interaction $\mathcal{J}$, coupled with a lattice structure featuring more neighboring sites, is associated with a higher critical temperature ($T_C$ or $T_N$). Additionally, the parameter $\Lambda$, which depends on $\mathcal{J}$, the lattice parameters ($a$ and $c$), and the spin quantum number ($\mathcal{S}$), is larger in materials with higher values of these parameters. For this reason, iron (Fe) exhibits the highest critical temperature and $\Lambda$ among the materials studied.

The maximum magnon specific heat ($C_m$), which peaks at the critical transition temperature, also depends on $\mathcal{J}$, $\mathcal{S}$, and whether the material is FM or AFM. CrSb, due to its strong exchange interactions and AFM ordering, exhibits the highest specific heat among the studied materials. In AFM materials, the degeneracy of magnons, their higher velocities, and generally longer magnon relaxation times contribute to larger $C_m$ values compared to FM materials. Notably, MnTe, despite its weaker exchange interactions, demonstrates a high specific heat, ranking third, close to that of Fe. The relatively high specific heat of MnTe, like that of CrSb, is attributed to its AFM nature, which tends to enhance $C_m$ compared to FM materials [34]. Figure 4 illustrates the specific heat values for comparison.*

The magnon contribution to the specific heat is determined by subtracting other contributions, such as electronic, phononic, and Schottky contributions, from the total specific heat. Although CrSb has the highest peak $C_m$, its specific heat per carrier is smaller than that of MnTe due to CrSb's larger carrier concentration ($n_e$). Generally, as $n_e$ increases, the magnon-drag thermopower ($\alpha_{md}$) decreases. This trend explains why metals with high $n_e$ exhibit lower $\alpha_{md}$ than semiconductors. Table 3 shows that semiconductors have lower carrier densities compared to the metals listed, and among them, MnTe has the lowest carrier density. Consequently, MnTe generates significantly larger entropy per carrier, resulting in a larger magnon-drag thermopower than the other materials.

MnTe's lower transition temperature also contributes to a larger entropy per carrier. As discussed earlier, specific heat ($C_m$) directly impacts magnetic entropy ($S_m$) through the relation $S_m = \int_0^T C_m/T'\, dT'$. A specific heat peak at a lower temperature increases the area under the $C_m/T$ curve, resulting in a higher $S_m$. Another good example is MnSb. In MnSb, despite a small $C_m$ value (Figure 4 (a)) and a steep decline after the critical temperature, the specific heat peak occurs at a lower temperature - just above that of MnTe and lower than the other materials. This low-temperature peak compensates for the small $C_m$, resulting in a $S_m/n_e$ value comparable to that of CrSb (Figure 4 (c)), which has a significantly higher $C_m$.

Furthermore, the slope of the $C_m$ curve's decrease after the critical temperature also affects $S_m$. The presence of paramagnons above the critical temperature causes this decrease to occur more gradually, as short-range magnetic order or paramagnons retain part of the magnetic specific heat at higher temperatures [29]. This gradual decline further enhances $S_m$ in MnTe.

MnTe stands out in terms of magnetic entropy per carrier, exhibiting values over two orders of magnitude higher than those of other materials, as shown in Figure 4(c). This is consistent with its significantly larger magnon-drag thermopower. Its remarkable magnon and paramagnon drag thermopower is attributed to a combination of properties discussed above, namely: **(i)** long magnon relaxation time ($\tau_m$), **(ii)** short electron relaxation time due to magnon scattering ($\tau_{em}$), **(iii)** large magnon group velocity, **(iv)** degenerate magnon modes, **(v)** large specific heat peak ($C_m$), **(vi)** low electron density ($n_e$), **(vii)** low Néel temperature ($T_N$), and **(viii)** gradual decline of $C_m$ above $T_N$. These factors collectively explain its notably large magnon and paramagnon drag thermopower.

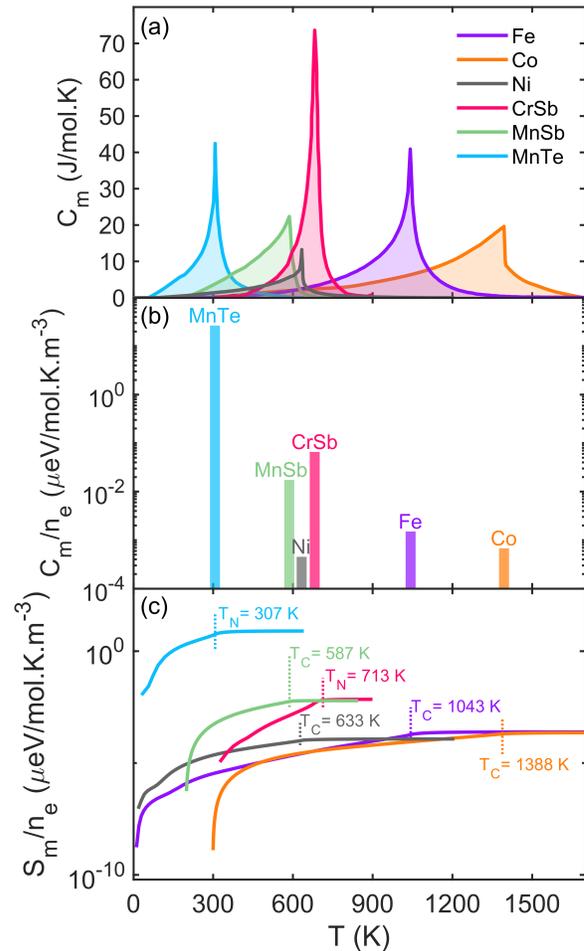

Figure 4: Comparison of magnon specific heat (a), magnon specific heat per carrier (b), and entropy per carrier (c) as a function of temperature for various magnetic materials Fe, Co, Ni, CrSb, MnSb, and MnTe.

---

* The data for Fe (Ref. [54]) and Ni (Ref. [55]) correspond to the magnetic contribution reported directly in the respective studies. For CrSb and MnSb, the total specific heat data were taken from Ref. [34], and the magnetic component was obtained by subtracting all other relevant contributions, such as phononic and electronic contributions. The specific heat data for Co and MnTe were obtained from our own experimental measurements.

## VI. CASE STUDY 2: THERMOPOWER OF SUPERCONDUCTORS NEAR $T_C$

Besides FM and AFM materials, superconductors show a sharp peak in electronic specific heat at the critical temperature due to the formation of Cooper pairs and the subsequent energy gap in the electronic density of states. For example, lead shows a sharp peak in specific heat at approximately 7.2 K, and niobium (Nb) exhibits a similar increase at 9.2 K. The specific heat of Nb, which is a type-I superconductor, can be observed in Figure 5. The total specific heat of this system comprises two terms: electronic and phononic. The phononic contribution is given by the Debye model, while the electronic contribution increases linearly above the transition temperature, as the system transitions from the superconducting to the normal phase. By considering these relationships and experimental data, the Debye temperature of 275K and a $\gamma$ value of 0.014 fits the experimental data within the experimental errors. The estimated phonon and electronic contributions can also be visualized in Figure 5.

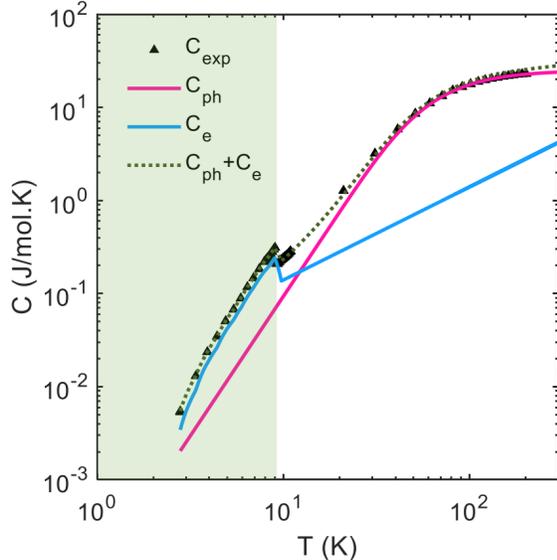

Figure 5: Specific heat of Nb as a function of temperature. Experimental data are shown as black triangles, while the phonon contribution is represented by the magenta line, and the electronic contribution by the blue line. The total specific heat, including both phonon and electronic contributions, is indicated by the dotted line.

Below the critical temperature, in the superconducting phase, the estimated electronic contribution increases non-linearly with temperature (because of the energy gap), shows a cusp when we approach the critical temperature and drops abruptly at the critical temperature. Above the critical temperature, in the normal phase the estimated electronic specific heat increases linearly with temperature, resulting in a temperature-dependent entropy which increases linearly by increasing temperature.

At temperatures well below the critical temperature, the electronic contribution to the specific heat of Nb follows an exponential dependence, $e^{-\Delta/k_B T}$, where $2\Delta$ is the superconducting gap, arising from the formation of Cooper pairs [56]. The estimated gap is approximately 2.58 meV, which is in reasonable agreement with values obtained from infrared absorption measurements [57].

To determine the number of electrons contributing to the electronic specific heat in one mole of Nb, we note its electron configuration [Kr] 4d⁴ 5s¹, indicating 5 valence electrons. In metals, it is primarily the valence electrons that contribute to the electronic specific heat, as these electrons can move relatively freely through the metal lattice and participate in conduction. Thus, approximately $5N_A$ electrons, where $N_A$ is Avogadro's number, are available to contribute to the electronic specific heat in one mole of Nb. However, the actual contribution of these electrons at a given temperature depends on the density of states near the Fermi level and the temperature itself, as not all electrons near the Fermi level will be thermally excited. In Figure 5, it is shown that $N = 1.6 N_A$ provides a better fit to the experimental data, reflecting the fact that only a fraction of the total valence electrons contributes significantly to the specific heat at the experimental range of temperature.

The thermopower plot in Figure 6 exhibits significant fluctuations below the superconducting transition temperature $T_C$, as well as irregularities extending slightly above $T_C$ up to approximately 14 K. These variations are likely due to signal-to-noise limitations, as the true thermopower is expected to be very small, on the order of or below 1 μV/K, in this temperature range. While theory predicts a rise in thermopower near $T_C$, the large fluctuations and poor signal fidelity in this regime prevent a conclusive observation of such behavior in the measured data. In fact, the presence of substantial fluctuations around zero is consistent with the expectation that the thermopower is nearly vanishing below $T_C$, making the fit highly sensitive to noise and baseline drift.*

---

* Data were measured for a Nb strip using a Quantum Design PPMS system. Below the superconducting transition temperature $T_C$, the true thermopower of Nb is expected to vanish, resulting in no thermoelectric voltage across the sample. However, the TTO module of the PPMS still applies a finite temperature gradient and fits the measured voltage response ΔV(t) to an empirical exponential model. In the absence of a real thermoelectric signal, this nonlinear fit is effectively applied to random noise and baseline drift. Because the model assumes the presence of an underlying physical trend, it often returns artificially large values for the steady-state voltage ΔV∞, leading to spurious thermopower values up to ±1 μV/K. Above $T_C$, even when the true thermopower is similarly small, the presence of a physically meaningful ΔV(t) results in a well-conditioned fit and much lower apparent noise in the computed thermopower.

We calculate the entropy and thermopower from the electronic specific heat ($C_e$), and the resulting data aligns well with the experimental thermopower measurements, providing strong experimental validation for our theoretical discussions. Interestingly, the specific heat does not consistently increase with temperature; instead, it exhibits a peak at the transition temperature ($T_C$) before transitioning to a linear rise. If thermopower were directly proportional to specific heat, it would be expected to follow a similar trend, decreasing immediately after $T_C$. However, the experimental data for niobium shows a different behavior, with thermopower not undergoing an immediate decrease post-transition.

This observation is consistent with the discussed theory, further reinforcing its applicability.

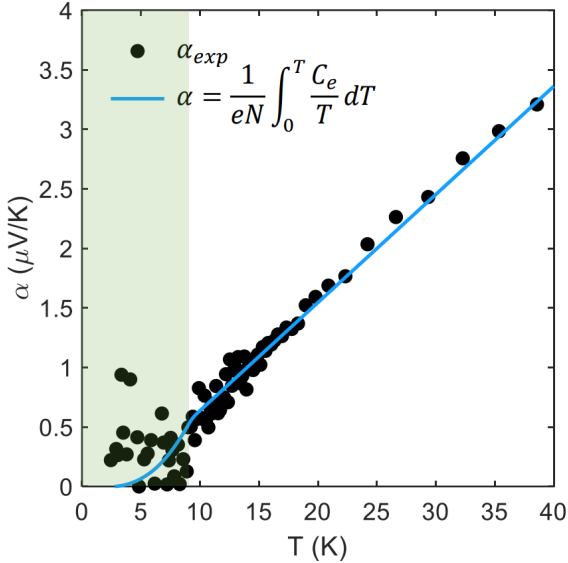

Figure 6: Thermopower ($\alpha$) of Nb as a function of temperature. Experimental data are shown as black circles, and the blue line represents the thermopower calculated using our theoretical framework. Significant fluctuations in the measured values appear below and near the superconducting transition temperature $T_C$, consistent with the expected vanishing thermopower and limitations in signal-to-noise ratio. Despite these experimental uncertainties, the overall agreement above $T_C$ supports the model's accuracy in capturing the thermoelectric behavior of Nb in the normal state.

## VII. CASE STUDY 3: THERMOPOWER AT NANOSCALE

A molecular junction consists of one or more organic or inorganic molecules, stretched between two macroscopic electrodes (see Figure 7). These systems are atomic-scale, significantly smaller than the electron mean free path in the electrodes [58]. Consequently, transport in molecular junctions necessitates a full quantum description, unlike bulk systems that can be analyzed using a semiclassical method [59].

As in bulk materials, when the two sides of a molecular junction are at different temperatures, $T_L$ and $T_R$, a voltage difference, $\phi_L - \phi_R = e(\mu_L - \mu_R)$, emerges. This voltage difference is proportional to the temperature gradient and is represented as:

$$\Delta\phi = \phi_L - \phi_R = -\alpha(T_L - T_R). \quad (46)$$

The flow of electrical current through a nanoscale object, such as a molecular junction, is determined by the transmission probability of an electron successfully transmitting through the junction from the left-hand side electrode to the right-hand side electrode.

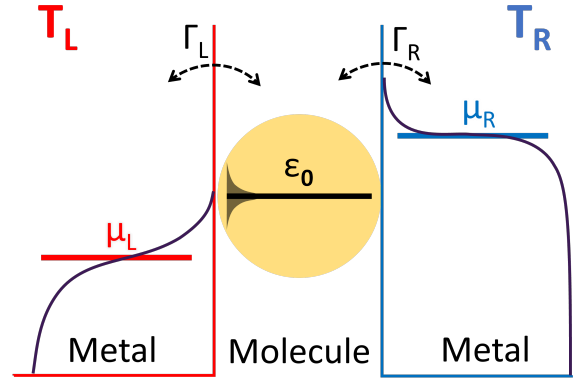

Figure 7: Schematic illustration of a single molecular state at energy $\varepsilon_0$ bridging two contact leads. When a temperature gradient is applied across the junction, depending on the position of the chemical potential in the molecule with respect to $\varepsilon_0$, the chemical potential on the hot side ($\mu_L$) shifts lower or higher than that on the cold side ($\mu_R$) to cancel net current flow. This shift generates a Seebeck voltage, bringing the system to equilibrium.

This transmission varies with the energy of the electron and is influenced by the molecule's electronic structure, particularly its molecular orbitals. These orbitals are further modified by coupling to the contacts. Thus, for systems where charge carriers do not interact, the current $I$ can be written using Landauer's formula as [60]:

$$I = \frac{2e}{h}\int_0^\infty d\varepsilon \mathcal{T}(\varepsilon)[N_L(\varepsilon) - N_R(\varepsilon)], \quad (47)$$

where,

$$\mathcal{T}(\varepsilon) = \frac{4\Gamma_L \Gamma_R}{(\varepsilon - \varepsilon_0)^2 + (\Gamma_L + \Gamma_R)^2}, \quad (48)$$

represents the transmission probability for an electron with energy $\varepsilon$. Here, $\varepsilon_0$ is the energy level of the molecule, $h$ is Planck's constant, $\Gamma_L$ and $\Gamma_R$ denote the coupling of the molecule to the left and right leads, respectively. $N_L$ and $N_R$ are the Fermi-Dirac distribution functions for the left and right electrodes, given by

$$N_i(\varepsilon) = \left(1 + e^{(\varepsilon-\mu_i)/k_B T_i}\right)^{-1}; \quad i = L, R,$$

where $\mu_i$ represents the chemical potential and $T_i$ denotes temperature of the corresponding electrode (see Figure 7). By defining the temperature and chemical potential of the left and right leads as:

$$T_L = T + \frac{\Delta T}{2}, \quad T_R = T - \frac{\Delta T}{2},$$
$$\mu_L = \mu + \frac{e\Delta\phi}{2}, \quad \mu_R = \mu - \frac{e\Delta\phi}{2} \quad (49)$$

where $\mu$ is the chemical potential and $T$ is the temperature of the molecule, and performing simple calculations, detailed in Appendix E, we can derive the thermopower as [60]:

$$\alpha(T) = \frac{1}{eT} \frac{\int_0^\infty \mathcal{T}(\varepsilon)(\varepsilon-\mu)\left(-\frac{\partial N}{\partial \varepsilon}\right) d\varepsilon}{\int_0^\infty \mathcal{T}(\varepsilon)\left(-\frac{\partial N}{\partial \varepsilon}\right) d\varepsilon}. \quad (50)$$

In general, $\mathcal{T}(\varepsilon)$ is energy-dependent. However, if we assume an energy-independent transmission, the expression for the thermopower can be simplified as in Eq. (10), similar to the thermopower of a bulk material.

The transmission probability described in Eq. (48), takes on a Lorentzian form. In the case of symmetric coupling, where $\Gamma_L = \Gamma_R = \Gamma/2$, the transmission function has a full width at half maximum of $2\Gamma$. As $\Gamma \to 0$, indicating negligible broadening due to coupling with the electrodes, the Lorentzian function approaches a delta function:

$$\mathcal{T}(\varepsilon) = \Gamma\delta(\varepsilon - \varepsilon_0). \quad (51)$$

Under this condition, the thermopower can be evaluated from Eq. (50), yielding:

$$\alpha(T) = \frac{k_B}{e}\left(\frac{\varepsilon_0 - \mu}{k_B T}\right). \quad (52)$$

As shown in Eq. (52), the thermopower depends explicitly on the energy level of the molecule ($\varepsilon_0$), the chemical potential ($\mu$), and the temperature ($T$). Notably, $\alpha$ vanishes when $\varepsilon_0 = \mu$, becomes negative when $\varepsilon_0 < \mu$, and takes positive values when $\varepsilon_0 > \mu$, reflecting the particle-hole asymmetry inherent in the system.

We now demonstrate that the thermopower is directly proportional to the system's entropy and can be expressed as the derivative of entropy with respect to the occupation number $N$ of the single level $\varepsilon_0$.

The entropy of the molecule is given in terms of the occupation number $N$ as:

$$S(T) = -k_B[N \ln N + (1-N)\ln(1-N)], \quad (53)$$

where $N = 1/(1 + \exp\left(\frac{\varepsilon_0 - \mu}{k_B T}\right))$. It is important to note that the occupation number of an isolated single molecular level remains fixed, resulting in zero entropy. However, when the molecule is coupled to external leads, charge exchange induces fluctuations in the occupation number. These fluctuations introduce additional accessible microstates, thereby increasing the system's entropy.

The specific heat of this system is given by:

$$C_V(T) = \frac{\varepsilon_0^2}{k_B T^2} N(1-N), \quad (54)$$

where $N(1-N)$ represents the variance of the occupation number associated with the energy level $\varepsilon_0$. In the absence of fluctuations - i.e., when $N$ is either 0 or 1 - this variance vanishes, leading to a zero specific heat.

Based on Eq. (17), differentiating the entropy with respect to $N$, yields:

$$\alpha(T) = -\frac{k_B}{e}[\ln N - \ln(1-N)]$$
$$= \frac{k_B}{e}\left(\frac{\varepsilon_0 - \mu}{k_B T}\right). \quad (55)$$

This result is consistent with Eq. (52), reinforcing our assertion that thermopower in open nanoscale systems can be directly derived from the entropy's dependence on particle number.

## VIII. CONCLUSION

This study has investigated the intricate relationship between thermopower, specific heat, and entropy across diverse systems, with a particular focus on magnon and magnon-drag thermopower in magnetic semiconductors and superconductors. Our findings demonstrate that thermopower is fundamentally linked to the degree of disorder within a system, which is inherently captured by entropy. We have shown that entropy serves as a natural variable for thermopower, akin to internal energy. In systems where specific heat exhibits a power-law dependence on temperature - such as Fermi gases and Fermi liquids - thermopower can be accurately described using specific heat. However, in systems with more complex behaviors, including magnetic materials above their phase transition temperature, entropy provides a more accurate measure of thermopower.

Our analysis of magnon-drag thermopower in magnetic systems reveals that thermopower is directly proportional to changes in magnon entropy relative to magnon density. At low temperatures, where magnons follow the Bloch law, magnon thermopower is proportional to the magnon specific heat. At the critical temperature, a phase transition occurs from a ferromagnetic (or antiferromagnetic) to a paramagnetic phase. Above this transition temperature, the magnon specific heat sharply declines, but entropy - is derived as the temperature integral of $C_m/T$ - remains significant. Consequently, the magnon and magnon-drag thermopower, which are entropy-dependent, do not decline. In such cases, specific heat cannot fully explain thermopower trends, and entropy offers a more precise explanation. Furthermore, we established that drag thermopower is influenced by both entropy and short-range magnetic correlations, with the ratio of magnon relaxation time ($\tau_m$) to electron-magnon interaction relaxation time ($\tau_{em}$) playing a key role in determining the fraction of spin excitation contributions to thermopower.

Three case studies were conducted to apply and validate our model: (1) magnetic materials, including FM metals (Fe, Co, Ni), an FM semimetal (MnSb), an AFM semimetal (CrSb), and an AFM semiconductor (MnTe); (2) the superconductor Nb; and (3) a single-molecule junction, treated as an open quantum system. In all cases, thermopower was shown to depend on entropy per carrier. In Nb, the thermopower is directly influenced by carrier entropy: below the critical temperature, Cooper pair entropy dominates, while above the critical temperature, free electron entropy becomes significant. For magnetic materials, the relationship is more nuanced, involving magnons in the ordered phase and paramagnons in the disordered phase. The interaction between these quasiparticles and carriers facilitates the transfer of entropy - and hence a fraction of the thermopower - from spin excitations to the carriers. This transfer process is largely governed by the ratio of magnon relaxation time to electron-magnon interaction relaxation time, which determines the magnitude of the magnon-drag contribution. In single-molecule junctions at the nanoscale, the transmission coefficient governs the flow of charge carriers into and out of the molecular level, thereby setting the steady-state occupation of the molecular state and determining the effective chemical potential. As demonstrated in our case study, the thermopower in such systems is directly influenced by the alignment between the molecular energy level and the chemical potential. Notably, the chemical potential controls both the thermopower and the system's entropy, such that the derivative of entropy with respect to the occupation number reproduces the thermopower, even in this open quantum regime.

**APPENDIX A: DERIVATION OF TABLE 2**

**EXAMPLES**

In this section, we examine the thermodynamic properties of some of the systems mentioned in Table 2.

**1. Ideal gas:**

The partition function for an ideal gas of $N$ non-interacting particles is given by [13]:

$$Z_N = \frac{1}{N!}\left(\frac{V}{\lambda_{th}^3}\right)^N, \quad (A1)$$

where $\lambda_{th} = \hbar/\sqrt{2\pi m k_B T}$ is the thermal wavelength, and $V$ is the volume of the system. The internal energy ($U$), the Helmholtz free energy ($F$), the specific heat ($C_V$), and the entropy ($S$) are, respectively, obtained as follows:

$$U = \frac{3}{2}Nk_B T, \quad (A2)$$

$$F = Nk_B T[\ln(n\lambda_{th}^3) - 1], \quad (A3)$$

$$C_V = \frac{3}{2}Nk_B, \quad (A4)$$

$$S = Nk_B\left[\frac{5}{2} - \ln(n\lambda_{th}^3)\right], \quad (A5)$$

$$\alpha = k_B\left[\frac{3}{2} - \ln(n\lambda_{th}^3)\right], \quad (A6)$$

where $n = N/V$. It is observed that the specific heat per particle is a universal constant, while the entropy and thermopower are temperature-dependent.

**2. Degenerate Fermi gas:** The distribution function is represented by [61]:

$$f_k(T,\bar{\mu}) = \frac{1}{e^{\left(\frac{\varepsilon(k)-\bar{\mu}}{k_B T}\right)} + 1}. \quad (A7)$$

The number of fermions is given by:

$$N(T,\bar{\mu}) = \sum_k f_k(T,\bar{\mu}) = \int_0^\infty \frac{\rho(\varepsilon)d\varepsilon}{e^{\left(\frac{\varepsilon-\bar{\mu}}{k_B T}\right)} + 1}$$

$$= \frac{V}{3\pi^2}\left(\frac{2m}{\hbar^2}\right)^{\frac{3}{2}}\bar{\mu}^{\frac{3}{2}}\left[1 + \frac{\pi^2}{8}\left(\frac{k_B T}{\bar{\mu}}\right)^2 + \cdots\right], \quad (A8)$$

where the density of state (containing the spin degeneracy factor ($2S+1$) is given by:

$$\rho(\varepsilon)d\varepsilon = \left(\frac{2m}{\hbar^2}\right)^{\frac{3}{2}}\frac{(2S+1)V\sqrt{\varepsilon}}{(2\pi)^2}d\varepsilon. \quad (A9)$$

The internal energy and the specific heat are respectively given by:

$$U = \int_0^\infty \frac{\varepsilon \rho(\varepsilon) d\varepsilon}{e^{(\frac{\varepsilon - \bar{\mu}}{k_B T})} + 1}$$

$$= \frac{3}{5} N \overline{\mu(0)} \left[ 1 + \frac{5\pi^2}{12} \left( \frac{k_B T}{\overline{\mu(0)}} \right)^2 + \cdots \right], \quad (A10)$$

$$C_V = \frac{3}{2} N k_B \left( \frac{\pi^2}{3} \frac{k_B T}{\overline{\mu(0)}} \right) + \mathcal{O}T^3, \quad (A11)$$

Given that $\overline{\mu(0)}|_{\phi=0} = \varepsilon_F$, we have:

$$C_V = \frac{m k_F}{3 \hbar^2} k_B^2 T + \mathcal{O}T^3. \quad (A12)$$

In this case, the entropy is equal to the specific heat. By using Eq. (17), one can calculate the thermopower of a Fermi gas:

$$\alpha = \frac{\pi^2}{6} \frac{k_B T}{\varepsilon_F} \frac{k_B}{e}. \quad (A13)$$

**3. Fermi liquid:** The specific heat of a Fermi liquid with effective mass $m^*$ is related to the specific heat of the Fermi gas by $m^*/m$, so the specific heat of the Fermi liquid is [3]:

$$C_V^* = \frac{m^* k_F}{3 \hbar^2} k_B^2 T. \quad (A14)$$

Similarly, we have for the thermopower:

$$\alpha = \frac{\pi^2}{6} \frac{k_B T}{\varepsilon_F^*} \frac{k_B}{e}; \varepsilon_F^* = \frac{\hbar^2 k_F^2}{2m^*}. \quad (A15)$$

In this case, the entropy is equal to the specific heat.

**4. High-temperature magnetic spin system:** At high temperatures, above the critical temperature, the long-range magnetic order in a spin system disappears, but short-range magnetic correlations persist. In this regime, magnons are quenched by thermal fluctuations, and the system exhibits paramagnetic behavior. The specific heat of the magnetic spin system follows a power-law decay, typically proportional to $\sim T^{-2}$. Despite the decline in specific heat, entropy remains significant because it is the integral of specific heat with respect to temperature. Therefore, magnon (or paramagnon) thermopower does not decline, and due to the presence of short-range magnetic correlations, the paramagnon carrier drag thermopower persists above the critical temperature. This effect is particularly pronounced near the phase transition, where the magnon thermopower enhances. For instance, in MnTe, the thermopower includes contributions from magnon-hole interactions, leading to an enhancement near the Néel temperature that continues above it without decline.

## APPENDIX B: MAXWELL'S RELATIONS IN THE GRAND CANONICAL ENSEMBLE

We use the grand canonical ensemble:

$$\langle \widehat{H} \rangle \equiv U = \phi_G + TS + \mu T. \quad (B1)$$

In constant $V$ and $T$, we have:

$$\left( \frac{\partial U}{\partial \mu} \right)_{T,V} = \left( \frac{\partial \phi_G}{\partial \mu} \right)_{T,V} + T \left( \frac{\partial S}{\partial \mu} \right)_{T,V}$$

$$+ \mu \left( \frac{\partial N}{\partial \mu} \right)_{T,V} + N. \quad (B2)$$

On the other hand:

$$dF = -PdV - SdT + \mu dN, \quad (B3)$$

$$\phi_G = U - TS - \mu N = F - \mu N, \quad (B4)$$

$$d\phi_G = dF - \mu dN - Nd\mu$$
$$= -SdT - PdV - Nd\mu, \quad (B5)$$

$$\Rightarrow \left( \frac{\partial \phi_G}{\partial \mu} \right)_{T,V} = -N. \quad (B6)$$

According to Eq. (B6), the Eq. (B2) can be rewritten as:

$$\left( \frac{\partial U}{\partial \mu} \right)_{T,V} = T \left( \frac{\partial S}{\partial \mu} \right)_{T,V} + \mu \left( \frac{\partial N}{\partial \mu} \right)_{T,V}. \quad (B7)$$

## APPENDIX C: CALCULATION OF THE MAGNON DENSITY OF STATE AND LIFETIME

This appendix outlines the methodology for estimating the magnon lifetime ($\tau_m$) for different materials, as summarized in Table 3.

**Density of States for Magnons**

The magnon energy $\mathcal{E}$ depends on the material's magnetic ordering. In AFM materials at low energies, $\mathcal{E}$ is linear with respect to $q$, while in FM materials, $\mathcal{E}$ has a quadratic dependence. These relationships allow us to calculate the magnon density of states $g(\mathcal{E})$ for each case.

**AFM Case**

For AFM materials:

$$\mathcal{E} = \hbar \omega = \hbar v_m q \rightarrow dq = \frac{1}{\hbar v_m} d\mathcal{E}, \quad (C1)$$

The volume element in $q$-space becomes:

$$d^3 q = 2 \times 4\pi q^2 dq = 8\pi \left( \frac{\mathcal{E}}{\hbar v_m} \right)^2 \frac{1}{\hbar v_m} d\mathcal{E}. \quad (C2)$$

Thus, the magnon density of states is:

$$g(\mathcal{E}) = \frac{8\pi}{(\hbar v_m)^3}\,\mathcal{E}^2. \tag{C3}$$

Here, the factor of 2 accounts for magnon degeneracy in AFMs.

**FM Case**

For FM materials:

$$\mathcal{E} = \hbar\omega = \hbar \mathcal{D} q^2 \rightarrow dq = \frac{1}{2\sqrt{\mathcal{E}\mathcal{D}\hbar}}d\mathcal{E}, \tag{C4}$$

The volume element in q-space becomes:

$$d^3q = 4\pi q^2 dq = 4\pi \frac{\mathcal{E}}{\hbar \mathcal{D}} \frac{1}{2\sqrt{\mathcal{E}\mathcal{D}\hbar}} d\mathcal{E}. \tag{C5}$$

Thus, the magnon density of states is:

$$g(\mathcal{E}) = \frac{2\pi}{(\hbar \mathcal{D})^{\frac{3}{2}}} \sqrt{\mathcal{E}}. \tag{C6}$$

## APPENDIX D: EXPERIMENTAL DATA OF NIOBIOM THERMOELECTRIC PROPERTIES

This appendix presents the experimental data collected for the thermoelectric properties of niobium error, including thermopower ($\alpha$), thermal conductivity ($\kappa$), electrical resistivity ($\rho_e$), and specific heat ($C$). All measurements were conducted using the Quantum Design DynaCool Physical Property Measurement System (PPMS) Thermal Transport Option (TTO) across the full temperature range to explore the behavior of niobium under different thermal conditions. Each data point is accompanied by error bars, representing uncertainties stemming from measurement techniques and instrumental limitations. These data provide comprehensive insights into the thermoelectric behavior of niobium in both the metallic and superconducting phases.

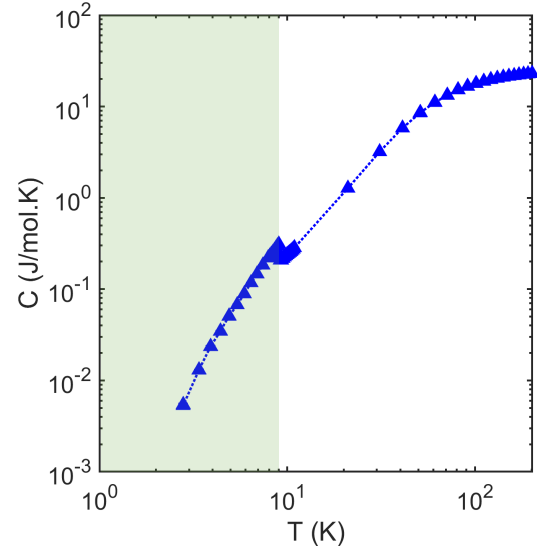

Figure D1: Specific heat ($C$) of Nb as a function of temperature, with error bars representing measurement precision (not visible as they are smaller than the symbols). The data show a pronounced peak at the critical temperature, corresponding to the phase transition in niobium.

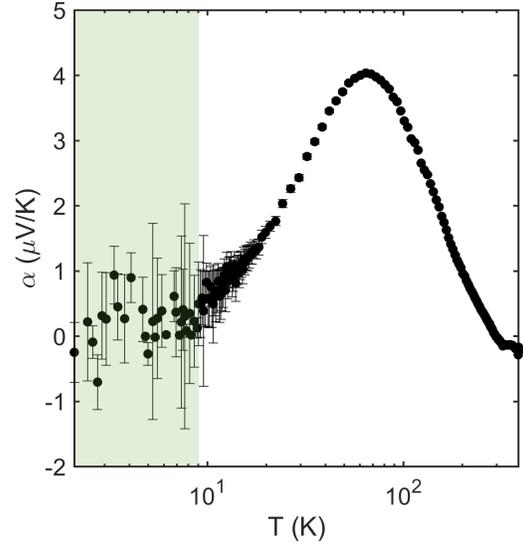

Figure D2: Thermopower ($\alpha$) as a function of temperature for Nb. Error bars represent the uncertainties in the measurement of the voltage difference and temperature gradient. Below $T_\text{C}$, where the thermopower is expected to be zero or near zero, the equipment (PPMS TTO option) fitting algorithm applies the model to small voltage signals. In this regime, minimal thermoelectric response leads to noise-dominated values, resulting in scattered, non-zero thermopower data. These fluctuations, evident in the plot, are artifacts of the fitting process and are not physically meaningful.

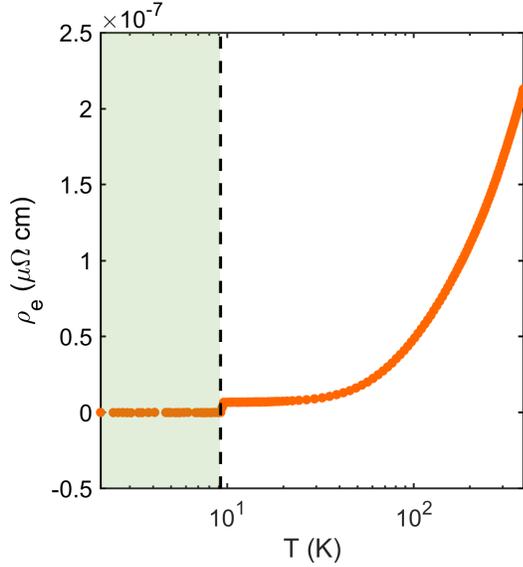

Figure D3: Temperature-dependent resistivity ($\rho_e$) of Nb. Error bars, representing uncertainties due to contact resistance and current measurement, are smaller than the symbols and therefore not visible. The plot highlights both metallic behavior above the critical temperature and superconducting behavior below it.

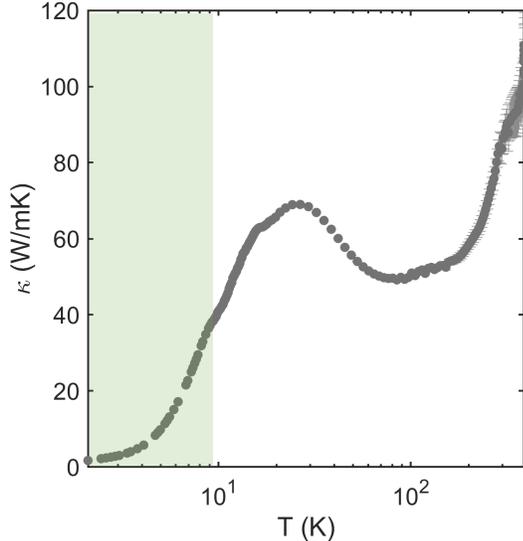

Figure D4: Thermal conductivity ($\kappa$) as a function of temperature. Error bars at lower temperatures are smaller than the symbols and therefore not visible.

## APPENDIX E: MOLECULAR JUNCTION THERMOPOWER

The difference between the left and right Fermi-Dirac distribution functions arises from both the temperature difference ($\Delta T$) and the chemical potential difference ($\Delta \mu$) between the leads given by:

$$N_L(\varepsilon) - N_R(\varepsilon) = \frac{1}{1 + e^{\frac{\varepsilon - (\mu - e\Delta\phi/2)}{k_B(T + \Delta T/2)}}} - \frac{1}{1 + e^{\frac{\varepsilon - (\mu + e\Delta\phi/2)}{k_B(T - \Delta T/2)}}}. \quad (E1)$$

To expand this expression to first order in $\Delta T$ and $\Delta \phi$, we use the Taylor series approximation for small perturbations around $T$ and $\mu$ for the left side:

$$N_L(\varepsilon, T_L, \mu_L) \approx N_L(\varepsilon, T, \mu) + \frac{\partial N}{\partial \mu}\bigg|_{T_0, \mu_0} \frac{\Delta \mu}{2} + \frac{\partial N}{\partial T}\bigg|_{T_0, \mu_0} \frac{\Delta T}{2}, \quad (E2)$$

and for the right side:

$$N_R(\varepsilon, T_R, \mu_R) \approx N_R(\varepsilon, T, \mu) - \frac{\partial N}{\partial \mu}\bigg|_{T_0, \mu_0} \frac{\Delta \mu}{2} - \frac{\partial N}{\partial T}\bigg|_{T_0, \mu_0} \frac{\Delta T}{2}, \quad (E3)$$

where $\Delta \phi = -\Delta \mu / e$. The derivative of the Fermi-Dirac distribution function is:

$$\frac{\partial N}{\partial \mu} = \frac{e^{(\varepsilon - \mu)/k_B T}}{k_B T(1 + e^{(\varepsilon - \mu)/k_B T})^2} = -\frac{\partial N}{\partial \varepsilon}. \quad (E4)$$

$$\frac{\partial N}{\partial T} = -\frac{(\varepsilon - \mu)e^{\frac{(\varepsilon - \mu)}{k_B}T}}{k_B T^2 \left(1 + e^{\frac{(\varepsilon - \mu)}{k_B T}}\right)^2} = -\frac{\partial N}{\partial \varepsilon}\left(\frac{\varepsilon - \mu}{T}\right). \quad (E5)$$

Applying a first-order Taylor expansion in $T$ and $\mu$:

$$N_L(\varepsilon) - N_R(\varepsilon) = -\frac{\partial N}{\partial \varepsilon}\Delta \mu - \frac{\partial N}{\partial \varepsilon}\left(\frac{\varepsilon - \mu}{T}\right)\Delta T. \quad (E6)$$

In equilibrium, $\Delta \phi = -\Delta \mu / e = -(\mu_L - \mu_R)/e$, and the current can be expressed as [60]:

$$I = G\Delta \phi + L_T \Delta T = G\Delta \phi + G\alpha \Delta T, \quad (E7)$$

where the conductance $G$ and $L_T = G\alpha$ are given by:

$$G = \frac{2e^2}{h}\int_0^\infty d\varepsilon \mathcal{T}(\varepsilon)\left(-\frac{\partial N}{\partial \varepsilon}\right), \quad (E8)$$

$$L_T = G\alpha = \frac{2e}{h}\frac{1}{T}\int_0^\infty d\varepsilon \mathcal{T}(\varepsilon)(\varepsilon - \mu)\left(-\frac{\partial N}{\partial \varepsilon}\right). \quad (E9)$$

Thus, the thermopower is explicitly given by:

$$\alpha(T) = \frac{1}{eT}\frac{\int_0^\infty \mathcal{T}(\varepsilon)(\varepsilon - \mu)\left(-\frac{\partial N}{\partial \varepsilon}\right)d\varepsilon}{\int_0^\infty \mathcal{T}(\varepsilon)\left(-\frac{\partial N}{\partial \varepsilon}\right)d\varepsilon}. \quad (E10)$$

**Acknowledgment:** This study is partially based upon work supported by the NSF under grant number CBET-2110603 and the AFOSR under contract number FA9550-23-1-0302.